\documentclass[traditabstract]{aa}
\usepackage{amsmath}
\usepackage{txfonts}
\usepackage{graphicx}
\usepackage{natbib}
\usepackage{amsfonts}
\usepackage{amssymb}
\usepackage{xcolor}


\bibpunct{(}{)}{;}{a}{}{,} 

\renewcommand{\hat}[1]{\oldhat{\mathbf{#1}}}

\def\JApA{\ref@jnl{JApA}}                   

\begin{document}

\title{Rotation and magnetism of \emph{Kepler} pulsating solar-like stars}
\subtitle{Towards asteroseismically calibrated age-rotation relations}
\titlerunning{Rotation and magnetism of \emph{Kepler} pulsating solar-like stars}
\author{R.~A. Garc\'\i a\inst{1} \and 
T. Ceillier \inst{1} \and 
 D.~Salabert \inst{1} \and
S. Mathur\inst{2} \and
J.~L. van Saders\inst{3} \and
M.~Pinsonneault \inst{3} \and
J. Ballot \inst{4, 5} \and
P. G. Beck \inst{1,6} \and
S. Bloemen \inst{7} \and
T.~L. Campante\inst{8} \and
G.~R. Davies \inst{1,8} \and
J.-D.~do Nascimento Jr.\inst{9,10}\and
S.~Mathis \inst{1} \and
T.~S. Metcalfe \inst{2, 11} \and
M.~B. Nielsen \inst{12,13} \and
 J.~C.~Su\'arez \inst{14}\and
W.~J. Chaplin\inst{8}
A.~Jim\'enez \inst{15, 16} \and
C.~Karoff \inst{11}
}
\institute{Laboratoire AIM, CEA/DSM -- CNRS - Univ. Paris Diderot -- IRFU/SAp, Centre de Saclay, 91191 Gif-sur-Yvette Cedex, France
\and Space Science Institute, 4750 Walnut Street, Suite 205, Boulder, CO 80301, USA
\and Department of Astronomy, The Ohio State University, Columbus, Ohio 43210, USA
\and CNRS, Institut de Recherche en Astrophysique et Plan\'etologie, 14 avenue Edouard Belin, 31400 Toulouse, France
\and Universit\'e de Toulouse, UPS-OMP, IRAP, 31400 Toulouse, France
\and Instituut voor Sterrenkunde, Katholieke Universiteit Leuven, Celestijnenlaan 200D, B-3001 Leuven, Belgium
\and Department of Astrophysics, IMAPP, Radboud University Nijmegen, PO Box 9010, NL-6500 GL Nijmegen, The Netherlands
\and School of Physics and Astronomy, University of Birmingham, Edgbaston, Birmingham B15 2TT, UK
\and Harvard-Smithsonian Center for Astrophysics, Cambridge, Massachusetts 02138, USA
\and U. Federal do Rio Grande do Norte, UFRN, Dep. de F\'{\i}sica Te\'orica e Experimental, DFTE, CP 1641, 59072-970, Natal, RN, Brazil
\and Stellar Astrophysics Centre, Dept. of Physics and Astronomy, Aarhus Univ., Ny Munkegade 120, DK-8000 Aarhus C, Denmark 
\and Institut f\"ur Astrophysik, Georg-August-Universit\"at G\"ottingen, Friedrich-Hund-Platz 1, 37077, G\"ottingen, Germany
\and Max Planck Institute for Solar System Research Justus-von-Liebig-Weg 3 37077 G\"ottingen, Germany
\and Instituto de Astrof\'\i sica de Andaluc\'\i a (CSIC), 3004, Granada, Spain
\and Uni\-ver\-si\-dad de La Laguna, Dpto de Astrof\'isica, 38206, Tenerife, Spain
\and Ins\-ti\-tu\-to de Astrof\'\i sica de Canarias, 38205, La Laguna, Tenerife, Spain
}
\date{\today}
\abstract{\emph{Kepler} ultra-high precision photometry of long and continuous observations provides a unique dataset in which surface rotation and variability can be studied for thousands of stars. Because many of these old field stars also have independently measured asteroseismic ages, measurements of rotation and activity are particularly interesting in the context of age-rotation-activity relations. In particular, age-rotation relations generally lack good calibrators at old ages, a problem that this \emph{Kepler} sample of old-field stars is uniquely suited to address. 
We study the surface rotation and photometric magnetic activity of a subset of 540 solar-like stars on the main-sequence and the subgiant branch for which stellar pulsations have been measured. 
The rotation period was determined by comparing the results from two different analysis methods: i) {the projection onto the frequency domain of the time-period analysis}, and ii) the autocorrelation function (ACF) of the light curves. Reliable surface rotation rates were then extracted by comparing the results from two different sets of calibrated data and from the two complementary analyses. General photometric levels of magnetic activity in this sample of stars were also extracted by using a photometric activity index, which takes into account the rotation period of the stars.
We report rotation periods for 310 out of 540 targets (excluding known binaries and candidate planet-host stars); our measurements span a range of 1 to 100 days\footnote{Tables 3 and 4 are only available in electronic form at the CDS via anonymous ftp to cdsarc.u-strasbg.fr (130.79.128.5) or via http://cdsweb.u-strasbg.fr/cgi-bin/qcat?J/A+A/}. The photometric magnetic activity levels of these stars were computed, and for 61.5\% of the dwarfs, this level is similar to the  range, from minimum to maximum, of the solar magnetic activity. We demonstrate that hot dwarfs, cool dwarfs, and subgiants have very different rotation-age relationships, highlighting the importance of separating out distinct populations when interpreting stellar rotation periods.  Our sample of cool dwarf stars with age and metallicity data
of the highest quality is consistent with gyrochronology relations reported in the literature.
}
\maketitle

\section{Introduction}

Stellar rotation fundamentally modifies stellar interiors \citep[e.g.][]{1992A&A...265..115Z, 1997ARA&A..35..557P, 2004A&A...425..229M,2009A&A...495..271D, 2010A&A...519A.116E, 2013A&A...555A..54C, 2013A&A...549A..74M}. When this is considered in the stellar evolution models, the inferred age of the star is modified, which has severe consequences
in planetary systems, for example  \citep[e.g.][]{2009Natur.462..168P}. Surface rotation can also be used as an observable to determine the age of the star. Gyrochronology --the empirical relationship between rotation period, color, and age -- provides means by which surface rotation can be used to  infer ages of cool stars. These relationships, however, must be calibrated for different stellar populations, and rely on systems in which both rotation periods and stellar ages can be independently measured. The relationship between rotation period and age was first noted by  \citet{1972ApJ...171..565S}, followed by many studies relating rotating periods to magnetic activity \citep[e.g.][]{1984ApJ...287..769N,1984ApJ...279..763N,1987A&A...177..155R,1995A&A...294..515H,2008ssma.book.....S}. In 1999, \citeauthor{lachaume1999} used surface rotation rates for the first time as an age diagnostic. Gyrochronology relations have since been developed and refined by works such as \citet{pace2004}, \citet{2007ApJ...669.1167B}, \citet{cardini2007}, \citet{mamajek2008}, \citet{2009ApJ...695..679M}, \citet{2011ApJ...733..115M}, \citet{barnes2010}, and \citet{2012ApJ...746..102C} and have been used to unveil solar twins and analogs in the \emph{Kepler} sample of stars \citep{2014ApJ...790L..23D}. In general, these authors found relationships with a similar time dependence, but expanded upon the color and mass dependencies that become evident with larger datasets. Corresponding theoretical work seeks to provide angular momentum loss laws that reproduce the observed spin-down \citep{1988ApJ...333..236K,reiners2012} and produce the gyrochronology relationship. Both these theoretical endeavours and a refinement of gyrochronology as a tool require independent period and age calibrators to make further progress. This existing body of work focuses exclusively on main-sequence stars, but subgiants and evolved stars are also expected to show an interesting period evolution \citep[e.g.][]{schrijver1993} with a different relationship between period and age \citep{2013ApJ...776...67V}.

During the past decades, the detailed knowledge of stellar evolution has been improved thanks to the observational constraints provided by helio- and asteroseismology \citep[e.g.][]{2010AN....331..866C}. Solar-type stars show global physical characteristics similar to the Sun. In particular, they have stochastically excited modes as a result of an outer convective envelope \citep[e.g.][]{1977ApJ...212..243G,1994ApJ...424..466G,2008A&A...478..163B}. 


Solar-type stars are generally slow rotators (in most cases $v \sin \rm{i} < 20$ kms$^{-1}$), and the influence of rotation on the oscillation frequencies is well known. However, distortion due to the centrifugal force can have a strong impact on the oscillation frequencies even for slow rotators \citep[e.g.][and references therein]{2009LNP...765...45G,2010AN....331.1038R}. This effect is stronger for acoustic (p) modes with small inertia, which are more sensitive to the outer layers of the star. Therefore, their frequencies are more sensitive to the physical properties of the surface, where the centrifugal force becomes more efficient \citep[e.g.][]{2010ApJ...721..537S,2012A&A...542A..99O}. The induced perturbations are of the same order of magnitude --or even stronger-- as the effects of turbulence or diffusion, which
are currently considered the origin of the so-called surface effects.
It is thus important to study the surface rotation of other stars that are similar to the Sun and have higher rotational velocities. This could allow us to better understand the role of rotation in comparison to other surface effects and, hence, to properly interpret the oscillation spectra of solar-like stars, at least in the high-frequency domain.

With the advent of the detection of mixed modes in subgiant and giant stars 
\citep[e.g.][]{2011Sci...332..205B,2011Natur.471..608B,2011A&A...532A..86M,2013ApJ...765L..41S,2013ApJ...767..158B,2014ApJ...781L..29B} --including some belonging to a few clusters \citep{2011ApJ...739...13S,2011ApJ...729L..10B}-- it is now possible to infer the core rotation rate of these stars \citep{2012Natur.481...55B,2012A&A...548A..10M,2013EPJWC..4303012D,2013A&A...549A..75G}, which is still difficult to determine for solar-like stars and even for the Sun \citep[e.g.][]{GarCor2004,2007Sci...316.1591G,2008AN....329..476G}. In a few cases it is not only possible to obtain an averaged internal rotation rate 
\citep[e.g.][]{2013PNAS..11013267G,2013A&A...549A..12M,2013ApJ...766..101C}, but indications of the rotation profile from the external outer convection zone to the inner radiative core using inversion techniques \citep[][]{2012ApJ...756...19D,2014A&A...564A..27D,2014A&A...564A..36B}, as is also commonly done for the Sun \citep[e.g.][]{ThoJCD2003,2008SoPh..251..119G,2008A&A...484..517M,2013SoPh..287...43E}. Unfortunately, in the stellar case --because only low-degree modes are observable-- the inversion is increasingly uncertain close to the surface and measurements of the surface rotation rate are required. A better knowledge of the rotation profiles would be fundamental to answering long-standing questions about stellar interiors, transport of angular momentum, and rotational mixing. Hence the seismic analysis and the complementary study of the surface rotation are crucial.

High-quality photometric time series obtained by the \emph{Kepler} mission \citep{2010Sci...327..977B,2010ApJ...713L..79K} can be used to study the surface rotation and magnetic activity of solar-like stars in which eigenmodes are measured. Indeed, when a star is active \citep[e.g.][]{2010Sci...329.1032G,2010IAUS..264..120L}, starspots periodically cross the visible stellar disk. This produces a modulation in the brightness of the star that can be measured \citep[e.g.][]{2009A&A...506..245M,2012ASPC..462..133G,2012A&A...548L...1D}. The time evolution of these fluctuations provides a measurement of the surface velocity at the latitudes of the spots \citep[e.g.][]{2009A&A...506...41G,2011A&A...530A..97B,2013MNRAS.432.1203M,2013A&A...557L..10N}, which can also lead to a determination of the surface differential rotation \citep[e.g.][]{2009A&A...506...51B,2009arXiv0908.2244M,2010A&A...518A..53M,2012A&A...543A.146F,2013A&A...557A..11R,2014arXiv1402.6691L}.

The photometric variability of stars observed by \emph{Kepler} \citep[e.g.][]{2010ApJ...713L.155B,2013ApJ...769...37B} can be related to the surface magnetism at the time scales associated with the rotation periods. Therefore, the amplitude of the photometric modulation in the light curve can be used as an indicator of the surface magnetic variability, as recently demonstrated for the Sun \citep[e.g.][]{2013JPhCS.440a2020G,2014JSWSC...4A..15M}, on long and short timescales. Using a variability metric directly obtained from the light curve, \citet{2014ApJ...783..123C} confirmed that amplitudes of solar-like oscillations are suppressed in stars with increased levels of surface magnetic activity.


In the present work, we study the surface rotation rate and the photometric magnetic variability in the subset of 540 \emph{Kepler} solar-like stars studied by \citet{2014ApJS..210....1C}, for which accurate fundamental global parameters such as radius, masses, and ages have been inferred from the combination of asteroseismic and photometric observations. This seismic stellar sample has a potential impact on the field of gyrochronology, in which stellar rotation periods are used as a proxy for age. The period-age relations are empirically calibrated with stellar systems for which independent ages and rotation periods are measured. Several calibrations are reported in the literature \citep{pace2004,2007ApJ...669.1167B, mamajek2008,  2009ApJ...695..679M, 2011ApJ...733L...9M}, but they consistently struggle to find a calibration set at old ages and long rotation periods. \emph{Kepler} light curves provide means to measure both the stellar age through asteroseismology, and the rotation periods through spot modulation for an old field star population, and as such represent an important contribution to the gyrochronological calibrators.

We describe the preparation of the \emph{Kepler} light curves in Sect.~\ref{Sec:obs}, extract precise rotation periods in Sect.~\ref{Sec:rot}, and study the projected photospheric magnetic-activity levels in Sect.~\ref{Sec:sph}. Finally, in Sect. 5, we explore the correlation between asteroseismic age, rotation, evolutionary state, and mass. We also discuss surface magnetic activity diagnostics as inferred from \emph{Kepler} light curves.


\section{Observations and data analysis}
\label{Sec:obs}
We used data collected by the NASA planet-hunter mission \emph{Kepler}. The satellite is placed in a 372.5-d Earth-trailing heliocentric orbit. To keep the solar panels and the radiators that cool down the focal plane correctly aligned with respect to the Sun, \emph{Kepler} performs 90$^{\circ}$ rolls along its axis every three months \citep{2010ApJ...713L.115H}, which produces discontinuities in the observations. The light curves are therefore divided into ``\emph{Kepler} quarters'', denoted by Qn, starting by Q0 --the initial 10-d-long commissioning run-- followed by a 34-d-long first quarter (Q1), and subsequent three-month-long quarters (Q2, Q3, etc.). We used data up to Q14 whenever available for the stars in the sample.


About 120 000 stars --located in the constellation of Cygnus and Lyra-- have been quasi-continuously monitored by \emph{Kepler} during the full mission with a cadence of 29.4244 min (called long cadence, LC, data). Of all these stars, 512 can be observed with a much faster cadence of 58.84876~s (short cadence, SC, data) at any given time. This was done primarily to obtain a more precise timing of planetary transits \citep{2010ApJ...713L.160G}. However, a subset of around 120 stars of these 512 was reserved for asteroseismic studies by the \emph{Kepler} Asteroseismic Science Consortium (KASC, {http://kasoc.phys.au.dk/}) to study main-sequence and subgiant solar-like pulsating stars. During these evolutionary stages, the p-mode oscillations are located above the Nyquist frequency associated with the sampling rate of the LC data (283.45 $\mu$Hz). Therefore, the p-mode oscillations in these stars can only be studied using SC data (with a Nyquist frequency of  8.496 mHz) \citep[e.g.][]{2011A&A...534A...6C,2011ApJ...733...95M,2012A&A...543A..54A,2012ApJ...748L..10M}.

We are here interested in extracting the stellar surface rotation for periods longer than one day. Thus, we need to have light curves corrected for any low-frequency instrumental perturbations and with all the quarters properly concatenated. Therefore, we used simple aperture photometry (SAP) time series \citep{ThompsonRel21} and corrected for outliers, jumps, and drifts following the procedures described in \citet{2011MNRAS.414L...6G}, usually denoted as KADACS (\emph{Kepler} Asteroseismic Data Analysis and Calibration Software) light curves. These data were high-pass filtered using two triangular smoothing functions with cut-off periods at 30 and 100 days. 
The latter produces noisier light curves than the 30-day cut-off
period. Therefore we used the smaller filter when the results were the same, and we flag  in Table~\ref{tbl_rot} the results obtained  with the 100-day-filter light curves. Moreover, to minimize the dependency of the correction software, we also used  \emph{Kepler} light curves processed using pre-data conditioning maximum a posteriori methods \citep[PDC-MAP, e.g.][]{ThompsonRel21}. Unfortunately, this methodology works on a quarter-by-quarter basis, which filters all periods close to 90 days, as well as, sometimes, everything longer than 20 days with a growing attenuation for periods longer than 3 days. Finally, it is important to mention that we avoided using the latest \emph{Kepler} data products: pre-data conditioning multi-scale maximum a posteriori corrected series (PDC-msMAP) because each quarter is high-pass filtered with a limit of about 21 days. An example of the application of the three different correction methodologies applied to the same star can be seen in Fig.~1 of \citet{2013ASPC..479..129G}.

These corrections are not perfect, therefore some problems can remain in the corrected light curves. To minimise these effects, we removed from the light curves the quarters that show
an anomalously high variance compared with that of their neighbours. To do so, we computed the variance of every quarter in every light curve and divided the resulting array by its median. Then, we  computed the difference in this ratio between each quarter its two neighbours. If the mean of these two differences was greater than a threshold -- empirically set to 0.9 -- , we removed the quarter from the light curve. This method was applied to the PDC-MAP and the KADACS light curves to produce the datasets used in this study.


\begin{figure*}[t]
\begin{center}
\includegraphics[width=0.80\textwidth]{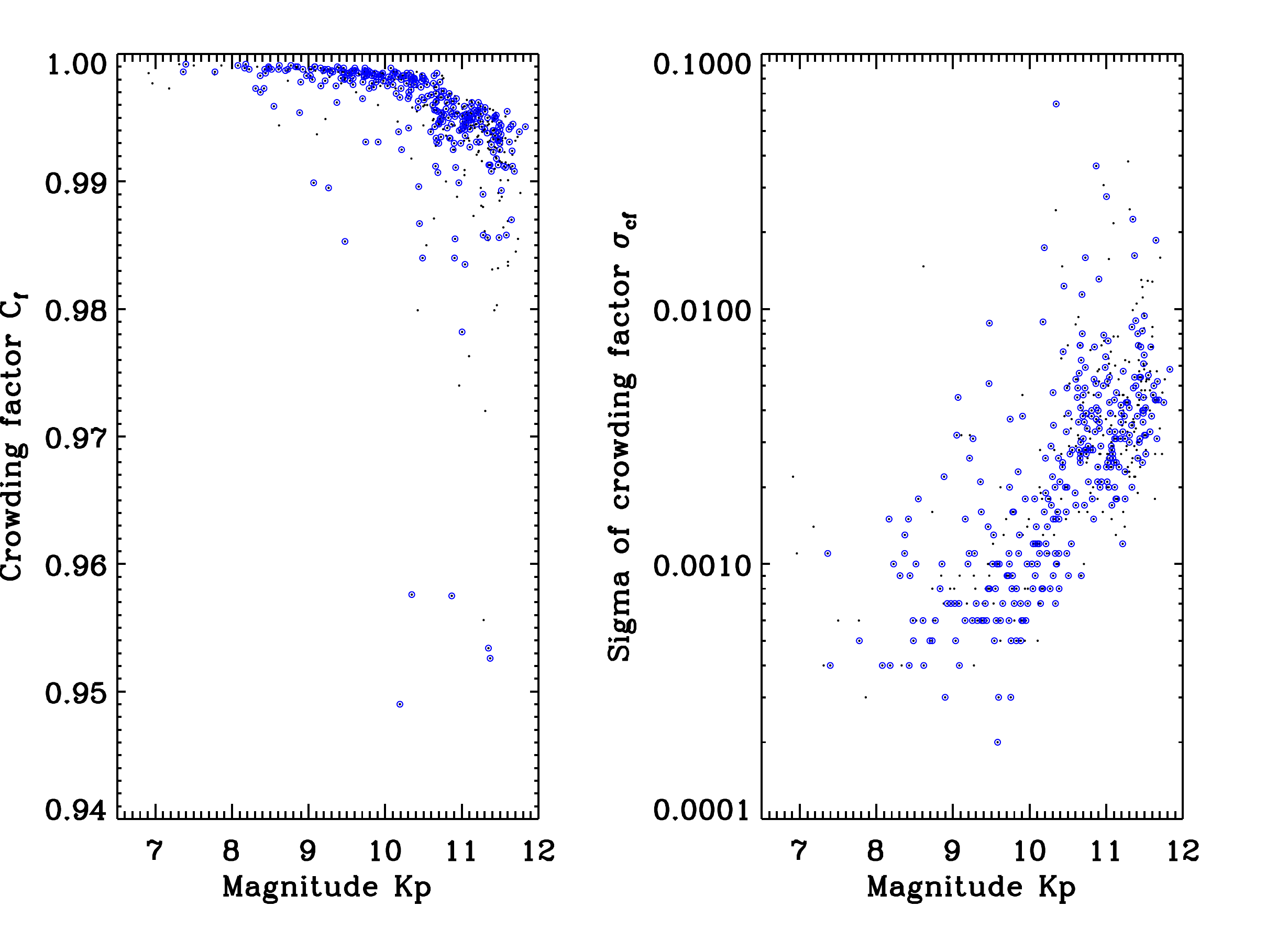}
\caption{Crowding factor $C_f$ (left panel) and sigma on the crowding factor $\sigma_{C_f}$ (right panel) for the 540 stars in the sample (black dots) as a function of the \emph{Kepler} magnitude. The 310 stars with measured $P_{\rm{rot}}$ are indicated by the blue circles.} 
\label{Fig:crowding}
\end{center}
\end{figure*}

Photometric pollution from nearby stars in the field of view can bias the rotation period estimate. The value of the crowding factor for each target (i.e. the ratio of the target-to-total flux in the optimal aperture) can be used to identify any target with a potential source of bias. Moreover, the crowding factor changes every quarter as \emph{Kepler} rolls along its axis. The median value of the crowding factor over all the observed quarters $C_f$ was therefore determined. The strongest variation of the crowding factor between each quarter, $\sigma_{C_f}$, was used as an indication of the associated uncertainty. The left panel of Fig.~\ref{Fig:crowding} shows the crowding $C_f$ values for all the 540 targets in the analysed sample as a function of the \emph{Kepler} magnitude $Kp$, while the right panel of Fig.~\ref{Fig:crowding} shows the associated uncertainties, $\sigma_{C_f}$. Most stars have a value of $C_f$ between 1 and 0.98, therefore we can postulate that they are not affected by pollution of another star in the field. Six of them appear as outliers with $C_f$ around 0.95. The pixel data of these stars were checked, validating their photometry. Only one star ($Kp$=11.0) lies beyond the represented y range: KIC~7938112, with a $C_f = 0.73 \pm 0.16$. The analysis of its power spectrum shows two humps of p-mode power, one corresponding to a solar-like star at high frequency, and another at low frequency corresponding to a red giant. Hence, this could be an example of a possible seismic binary or a case of a contaminated flux by a nearby red giant.

\begin{figure*}[!htb]
\begin{center}
\includegraphics[width=0.99\textwidth,height=0.77\textheight]{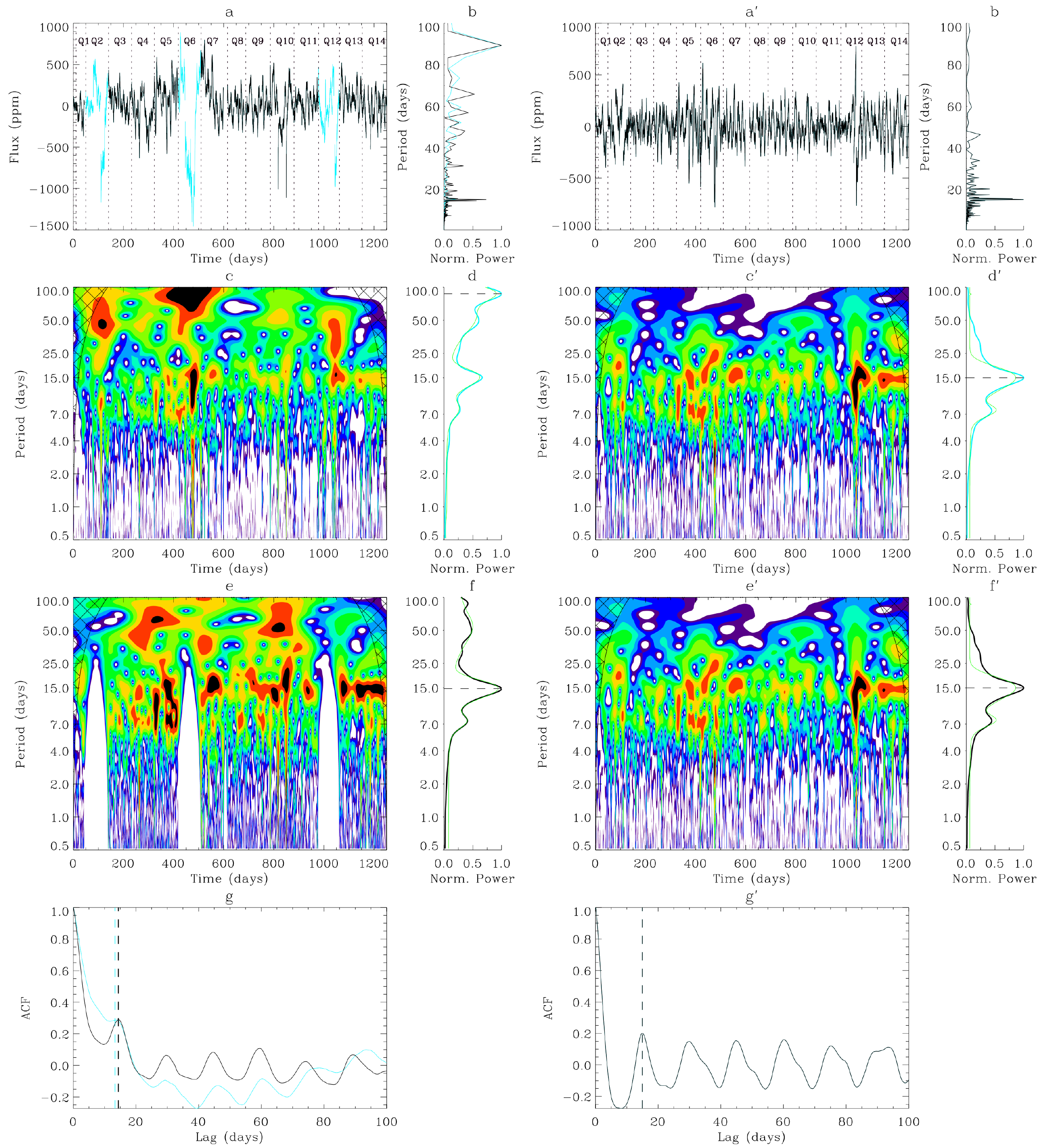}
\caption{Comparison of the two analyses used in this work, ACF and GWPS, applied on the two different datasets used Ð PDC-MAP (left-hand side panels) and KADACS (right-hand side panels marked with a prime) Ð for KIC~12258514. For each dataset, the plots are as follows: The top panels show (a) long-cadence \emph{Kepler} light curves (cyan) and the quarters selected for the analysis (black), where vertical dotted lines indicate the transitions between the observing quarters. The top right panels show (b)
the associated power density spectrum as a function of period between 0.5 and 100 days. The middle left panels (c and e) depict
the wavelet power spectrum (WPS) computed using a Morlet wavelet between 0.5 and 100 days on a logarithmic scale. The (c) panels correspond to the analysis of the entire light curve, while in the (e) panels only the selected portions of the light curve are used in the analysis. The black-crossed area is the cone of influence corresponding to the unreliable results. The middle right panels (d and f) plot the global wavelet power spectrum (GWPS) as a function of the period of the wavelet (thick cyan line for (d), thick black line for (f)) and the associated fit composed from several Gaussian functions (thin green lines). The horizontal dashed line designates the position of the retrieved $P_{\rm{rot}}$. Finally, the bottom panels (g) show the autocorrelation function (ACF) of the full light curve plotted between 0 and 100 days (cyan) and using only the selected portions of the light curve (black). The vertical dashed line indicates the returned $P_{\rm{rot}}$ for the ACF analysis.}
\label{Fig:1}
\end{center}
\end{figure*}

\section{Analysing the surface rotation}
\label{Sec:rot}
Several methods can be used to determine the average  rotation rate of the stellar surface that is caused by the motion of star
spots across the visible stellar disk. This periodicity can be measured by extracting the highest peak in the low-frequency part of the periodogram, as done by \citet{2013A&A...557L..10N}.  They studied thousands of dwarfs (from F- to M-type stars) observed by the {\it Kepler} mission during eight consecutive quarters, but analysed each one independently, which limited their analysis to periods in the range from 1 to 30 days. \citet{2013A&A...560A...4R} also studied the low-frequency part of the periodogram --computed using only Q3 data-- to extract rotation periods and differential rotation for almost 20 000 targets in a wider range of periods from 0.5 to 45 days. Unfortunately, the highest peak in the periodogram can sometimes be a higher overtone of the fundamental rotation period and the extracted rotation rate would be a multiple of the true value  \citep[see for further details][]{2013MNRAS.432.1203M}. To minimise this problem, we computed the wavelet power spectrum of the light curves and projected it onto the period axis, which reinforces the power of the fundamental peak and reduces the heights of the overtones \citep[for further details about the methodology see][]{2014A&A...562A.124M}.  \citet{2013MNRAS.432.1203M,2013ApJ...775L..11M,2014arXiv1402.5694M} also studied the autocorrelation function of the \emph{Kepler}  time series to determine rotation rates of M dwarfs, \emph{Kepler} Objects of Interests (KOI), and  about 34 000 main-sequence stars using PDC-MAP data from Q1 to Q4  for the first study and from Q3 to Q14 for the last two. By using the autocorrelation function, they demonstrated that they were less sensitive to instrumental problems than the direct analysis of the low-frequency part of the power spectrum.

In the present study we apply two of these methods: the autocorrelation of the light curves and the study of the low-frequency part of a time-period analysis projected onto the period domain (see next subsections for more details). We also use the time-period analysis to determine whether or not the rotation period detected is due to a localised instrumental glitch in the light curve or if it is a persistent feature in time that could then be interpreted as the true stellar rotation. All these analyses are performed using PDC-MAP and KADACS time series, rebinned by a factor of four to speed up the process, and put on a regularly spaced time array (see an example in the top panels of Fig~\ref{Fig:1}, corresponding to KIC~12258514). The comparison of the results obtained from different methods allows us to check the reliability of the extracted rotation period. Moreover, while most previous works analysed a single quarter or a few of them concatenated, we take advantage of much longer light curves from Q0 up to Q14.


\subsection{Time-frequency analysis and projected power spectrum}

The first study we performed to measure the surface rotation was a time-period analysis using a wavelet decomposition \citep{1998BAMS...79...61T}, improved for the low-frequency region as in \citet{Liu2007}, and adapted to our asteroseismic purposes following \citet{2010A&A...511A..46M}. This study allows us to track the temporal evolution of any modulation in the light curve that could be related to the rotation period and hence check whether it is not caused by a sudden event in the time series (normally related to a instrumental perturbation).


We used a Morlet wavelet -- which is the convolution of a sinusoid and a Gaussian function-- as the mother wavelet \citep{Goupillaud84cycle,Holds89}. The principle of the wavelet analysis consists of computing the correlation between the mother wavelet and the data by sliding the wavelet along the time axis of the light curves for a given scale or frequency of the wavelet. Then, we probed a range of scales for the wavelet and repeated the analysis. This produced the wavelet power spectrum (WPS) (see for example Fig.~\ref{Fig:1}, middle left panels). The red areas correspond to high power, while blue represents low power. This enabled us to determine
whether there was a rotation signature in the entire light curve. To increase our confidence in the rotation period estimate, we required that at least four rotations were observed in our light curves. This was delimited by the cone of influence that also takes into account edge effects (black-crossed area in the middle left panels of Fig.~\ref{Fig:1}).

Finally, we projected the WPS along the period axis yielding the global wavelet power spectrum (GWPS, thick black line in the middle right panels of Fig.~\ref{Fig:1}), which is similar to a Fourier power spectrum but with a degraded resolution. When the rotation period is well defined, the GWPS shows an almost perfect Gaussian profile. Therefore, to estimate the rotation period from the GWPS, we proceded as follows: In a first step, all the $N$ peaks in the GWPS with periods between 0.5 and 100 days were found.
Then, the GWPS was described and least-square minimised using the combinations from N to 1 Gaussian functions, removing iteratively the lowest peak. Finally, the fitted profile with the lowest reduced chi-squared was returned (green line in the the middle right panels of Fig.~\ref{Fig:1}).

The rotation period, $P_{\rm{rot}}$, of the star was first assumed to correspond to the highest fitted peak in the GWPS. An estimate of the uncertainty --including any possible differential rotation-- is given by the half width at half maximum (HWHM) of the fitted profile. 

Other harmonics of the signal can also be identified and fitted. In this way, one can verify if there are peaks with lower amplitudes that are multiples of the assumed $P_{\rm{rot}}$, in particular,  if there are multiples at higher periods that could indicate that the selected $P_{\rm{rot}}$ is the second or the third harmonic of the stellar rotation period.

\subsection{Autocorrelation function of the light curve}

The second analysis performed to derive $P_{\rm{rot}}$ was the autocorrelation of the light curves. This was done following a modified version of the procedure described by \citet{2013MNRAS.432.1203M}.  For a given light curve of a total duration $T$ and a time step $\delta t$, we computed the autocorrelation function, ACF, from $0$ to $0.5*T$ with a step $\delta t$. We then computed the power spectrum of the ACF to derive the most relevant period present in the ACF. Knowing this value, we smoothed the ACF with a Gaussian function of width a tenth of the selected period. This was done to minimise high-frequency variations in the ACF caused by noise and other high-frequency effects such as stellar pulsations.

After smoothing the ACF, we identified the first ten maxima whose heights were above a given threshold that was empirically set to 0.1. The first of these maxima --with the shortest period-- was assumed to correspond to $P_{\rm{rot}}$. In addition, we checked for the signature of a double dip or triple dip, that
is, for a structure in which one or two low peaks are followed by a higher peak. This shape of the ACF can be caused by several active regions on the star at different longitudes \citep[e.g. Fig.~2 in][]{2013MNRAS.432.1203M}. If such a structure is revealed to repeat itself, we selected as $P_{\rm{rot}}$ the highest  peak of the repeated sequence in the ACF. If we did not detect any peak above the threshold of 0.1, the star was assumed to be magnetically inactive (or having a very low inclination angle) and no $P_{\rm{rot}}$ was given. The bottom panels of Fig.~\ref{Fig:1} show an example of the ACF (black) and the smoothed version of the ACF (blue), which in this particular case are indistinguishable. The peak at about 15 days --marked by the vertical dashed line-- corresponds to the extracted rotation period. 

Even if the PDC-MAP light curve suffers from a noticeable jump and other instrumental instabilities, the autocorrelation function was able to extract the correct rotation period for both datasets, while the analysis of the GWPS failed when PDC-MAP data were used.


\subsection{Extracting stellar rotation periods}

While these methods are a relatively reliable means to derive the rotation period $P_{\rm{rot}}$, it is important to remember that the \emph{Kepler} light curves are divided into $\sim$90-day-long quarters and that the data are interrupted every month when the satellite faces Earth to download the recorded data. Thus, these two periodicities ($\sim$30 and $\sim$90 days) often appear in the \emph{Kepler} light curves and possibly in the results of the autocorrelation function analysis \citep{2014A&A...568A..10G}. For this reason, a comparison between the two sets of light curves and a visual check of the results are needed to ascertain the surface rotation period derived.

From the four values obtained using the two sets of data, PDC-MAP and KADACS, and the two analysis methods, GWPS and ACF, we extracted a reliable surface rotation period. To do so, we compared the four values and check whether they agreed within 20\%. If at least two results from two different sets agreed within this value, we considered this period to be the stellar rotation period. We defined $P_{\rm{rot}}$ and its uncertainty as the centre and the HWHM of the fit of the corresponding peak in the GWPS of the KADACS data set. We chose this solution because it is difficult to derive a reliable uncertainty on the rotation period from the ACF method, as pointed out by \citet{2013MNRAS.432.1203M}, and because we have a complete knowledge of all the corrections of the KADACS light curves.

After this first automatic comparison, we visually checked every light curve to verify that problems in the data did not prevent the automatic estimation of $P_{\rm{rot}}$. For some stars, the returned $P_{\rm{rot}}$ clearly comes from a common artefact in the two data sets, in which case we classified this period as unreliable. For some other stars, the light curves clearly showed a rotational modulation, but a glitch in one of the two sets prevented us from comparing the four results. In this case, we added these stars to the list of rotators, with a $P_{\rm{rot}}$ coming from the fit of the GWPS of the most stable of the two light curves, as indicated in Table~\ref{tbl_rot}. Finally, we separated the stars that show absolutely no rotation modulation from the stars that were just too noisy or unstable to obtain a reliable result. During this visual inspection we also found the stars for which a filter limit of 30 days was too short and re-filtered them with a filter limit of 100 days for the KADACS set before re-applying the same methodology to the new light curves.



\begin{table}
  \begin{center}
    \caption{\label{tbl_sample} Number of stars belonging to the different categories.}
    \begin{tabular}{lccc}
      \hline \hline
      Category & Total & Binaries & KOI \\
      \hline
      Whole sample & 540 & 15 & 15 \\
      \hline
      Period detected   & 321 & 11 & 12 \\
      No sign of rotation & 6 & 0 & 1 \\
      No reliable $P_{\rm{rot}}$ & 213 & 4 & 2\\
      \hline
    \end{tabular}
  \end{center}
\end{table}

\begin{figure}[!htbp]
\begin{center}
\includegraphics[width=0.5\textwidth]{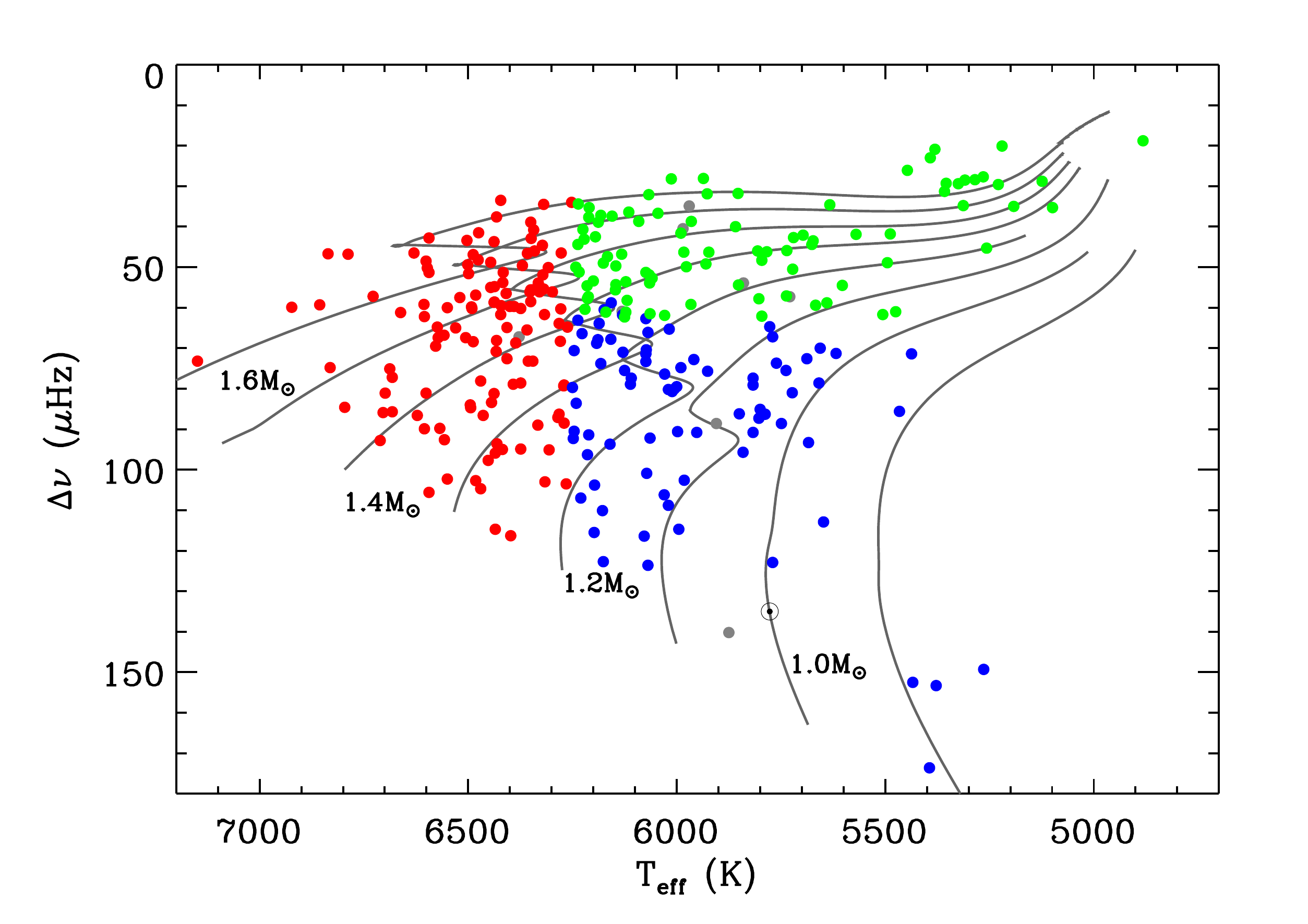}\\
\caption{Modified H.-R. diagram ($\Delta \nu$ vs $T_{\rm eff}$) showing the Sun and the 297 stars for which the rotation period, $P_{\rm{rot}}$, was successfully measured and a large frequency spacing is available from \citet{2014ApJS..210....1C}.  Hot stars are shown in red and defined as having $T_{\rm{eff}} >$6250. Dwarfs ($T_{\rm{eff}} \leq$ 6250 and $\log g > 4.0$) are shown in blue, and subgiants ($T_{\rm{eff}} \leq$ 6250 and $\log g \leq 4.0$) in green. Effective temperatures are taken from \citet{2011arXiv1110.4456P}. The stars for which only the effective temperatures from \citet{2014ApJS..211....2H} are available are plotted in grey. Evolution tracks, computed with the code ASTEC \citep{2008Ap&SS.316...13C}, are shown for a range of masses at solar composition (Z$_\odot$ = 0.0246).}
\label{Fig:HR}
\end{center}
\end{figure}

From the 540 stars of our sample, we found 321 for which we were
able to determine a reliable surface modulation period. Accordingly,
our sample can be divided into four different groups. The first group is composed of the 310 normal stars (i.e. non-binaries) for which we derived a surface rotation period (see Table~\ref{tbl_rot}). The second group is composed of the 15 binaries or multiple stars in our sample. These 15 stars are either eclipsing binaries from the  \emph{Kepler} Eclipsing Binary Catalog (http://keplerebs.villanova.edu, Kirk et al 2013, in preparation) for which our method returns the orbital period, double-lined spectroscopic binaries from \citep{2007Obs...127..313G} and \citet{2014arXiv1402.3794T}, stars flagged as \emph{Double or multiple star} or \emph{Star in double system} in the SIMBAD database (http://simbad.u-strasbg.fr/simbad/), or the previously mentioned possible seismic binary or star polluted by a nearby red giant -- KIC~7938112 -- for which our method detects low-frequencies solar-like oscillations. The results for these multiple stars are regrouped in Table~\ref{tbl_mult}. We also found 6 stars without signs of rotation in their light
curve (third group, see Table~\ref{tbl_no_rot}) and 213 stars whose light curves are too noisy to derive a reliable result (fourth group). This repartition is summarised in Table~\ref{tbl_sample}. Amongst the stars listed in Table~\ref{tbl_rot} are 12 KOIs. For each of these stars, we checked that the returned rotation period is not caused by the transiting planet candidate. In Fig.~\ref{Fig:HR} we show all stars for which a $P_{\rm{rot}}$ was derived and a large spacing is available from \citet{2014ApJS..210....1C}.  

These stars were divided into three groups based on the stellar parameters in \citet{2014ApJS..210....1C}: cool main-sequence dwarfs (blue, $T_{\rm{eff}} \leq$ 6250 K, $\log g >$ 4.0), hot stars (red, $T_{\rm{eff}} >$ 6250 K), and subgiants (green, $T_{\rm{eff}} \leq$ 6250K, $\log g \leq$ 4.0). To have a homogeneous set of temperatures we used those derived by \citet{2011arXiv1110.4456P},  who recalibrated the KIC photometry in the SDSS {\it griz} filters using YREC models.

\begin{figure}[!htbp]
  \begin{center}
    \includegraphics[width=0.51\textwidth]{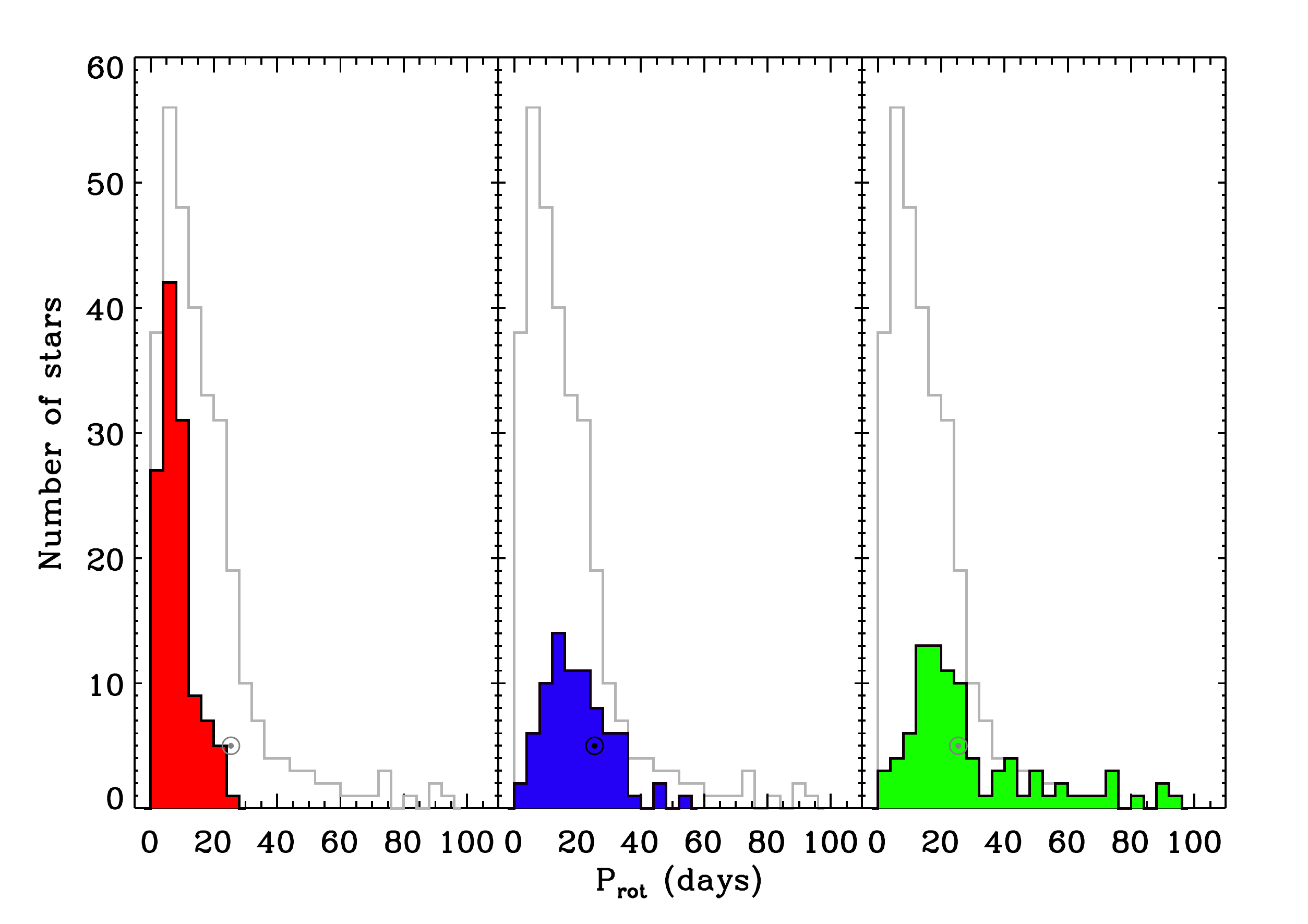}\\
    \caption{Histograms of the extracted surface rotation periods, $P_{\rm{rot}}$, for the full sample (grey), hot  (red), dwarf (blue), and subgiant stars (green) as defined in Fig.~\ref{Fig:HR}. For comparison with our sample of dwarfs, the solar rotational rate (25.4 days) is represented in black in the central panel at an arbitrary Y axis of 5. In the left and right panels, the Sun is plotted in grey as guidance for the eyes only.
    }
    \label{Fig:histo2}
  \end{center}
\end{figure}

The repartition of reliable $P_{\rm{rot}}$ can be seen in the form of a histogram in Fig~\ref{Fig:histo2}.  As we discuss in Sect.~\ref{sec:gyro}, hot stars rotate in general faster than cool main-sequence dwarfs, while the expansion of stars as they evolve on the subgiant branch leads to long rotation periods.


\section{Extracting the photometric magnetic activity level}
\label{Sec:sph}



Using CoRoT observations of HD~49933, \citet{2010Sci...329.1032G} demonstrated that  spots and magnetic features on the surface of a star follow the internal magnetic activity changes deduced from the analysis of the acoustic modes. A general index of stellar photometric variability, $S_{ph}$, was defined from the standard deviation of the whole light curve. In the case of the Sun, the comparison of $S_{ph}$ with a well-established magnetic activity proxy, the 10.7 cm radio flux (which is a useful proxy for the combination of chromospheric, transition region, and coronal solar EUV emissions modulated by bright solar active regions \citep[see for further details][]{2014JApA...35....1B}), demonstrated that $S_{ph}$ is a good indicator of the surface magnetic activity of the Sun \citep{2013JPhCS.440a2020G} and is well correlated with the chromospheric activity. \citet{2011AJ....141...20B} also defined a photometric variability index, called the {\it range}, $R_{var}(t_{len})$, to characterize the variability of the {\it Kepler} targets at different time scales.  This index, calculated by taking the flux included between 5\% and 95\% of the span in brightness, can underestimate the variability level of very active stars, however. Basri et al. (2013) calculated $R_{var}$ for the exoplanet targets Q9 time series of about 90 days, which were reduced using the PDC-MAP pipeline. $R_{var}$ was determined as the median value of segments of a given length, $t_{len} = 30$~days. This length was chosen because it is close to the solar rotation period.
However, the variability in the light curves can have different origins such as stellar pulsations, convection, or spots on the surface of the star, which are linked to the rotation period. For these reasons, and to specifically study stellar magnetic activity, the rotation period of the star needs to be taken into account in calculating the magnetic activity index. In this way, most of the variability should be related to the magnetism (spots) and not to the other sources of variability. However, it should be noted that the stellar inclination along the line of sight affects the observed value of the variability index if we assume that the stellar variability in solar-type stars is dominated by contributions from active latitudes as for the Sun. Consequently, an intrinsically active star observed at a low angle of inclination may present a moderate-to-low variability index.

Helioseismology has  proven that the surface magnetic activity is related to an inner dynamo process linked to the turbulence and the differential rotation between the envelope and the base of the convective zone.
The rotation period is thus a key parameter for understanding stellar magnetism. Moreover, when defining a stellar magnetic variability index for a large sample of stars, any temporal variations of the activity need to be taken into account, with periods of lower and higher activity, given the long time-series provided by the {\it Kepler} mission. The stellar variability indices determined so far \citep{2010Sci...329.1032G,2011AJ....141...20B,2013ApJ...769...37B} did not use the rotational period as input. In this section, we aim to determine a global magnetic activity index linked to the rotation period, which can also provide the possibility of studying the temporal evolution of the stellar activity. To do so, the Q0 to Q14 {\it Kepler} light curves for a total of over 1200~days were divided into sub-series of $k \times P_{rot}$, where $P_{\rm{rot}}$ is the rotational period of a given star, as measured in Sect.~\ref{Sec:rot}, and $k$ is an integer. For each individual sub-series, the standard deviation $S_{ph,k}$ of the non-zero values was calculated. The non-zero values exclude any quarter that was not observed for a given star or was removed because it was identified as {\it bad} (see Sec.~\ref{Sec:obs}). Moreover, a given sub-series was used only if the length of the sub-series was at least 2.5 times $P_{\rm{rot}}$  to avoid introducing any bias between sub-series. The magnitude correction of the photon noise from \citet{2010ApJ...713L.120J} was then applied to the $S_{ph,k}$. 

\begin{figure}[!htbt]
\begin{center}
\includegraphics[width=0.49\textwidth]{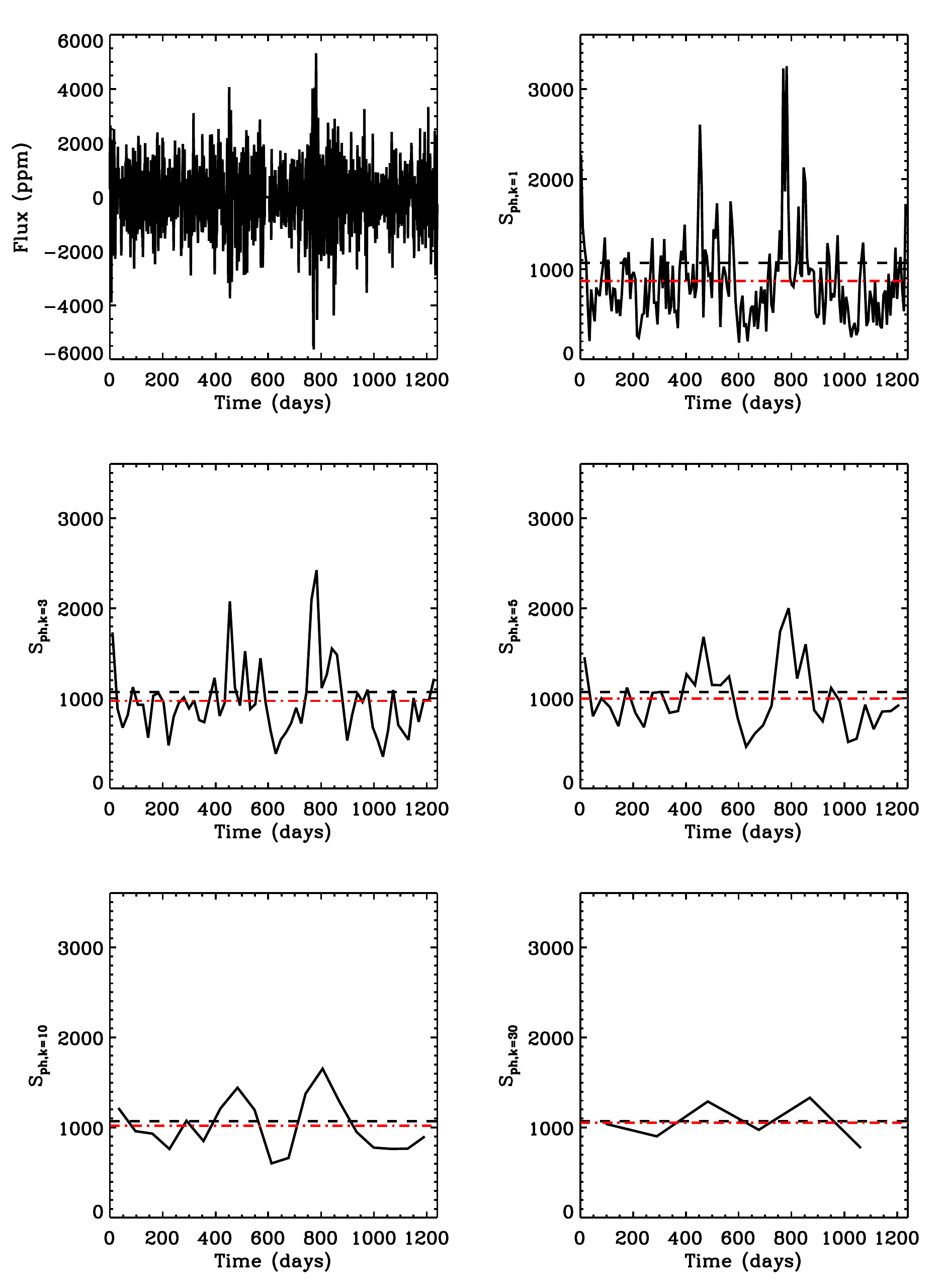}\\
\caption{Light curve of the solar-like pulsating star KIC~6448798 observed by {\it Kepler} from Q0 to Q14 (top left panel). From top right to bottom right panel, we show the photometric magnetic activity index, $S_{ph,k}$,  calculated using sub-series of size $k \times P_{\rm{rot}}$ with $k=1, 3, 5, 10$, and 30. The black dashed line represents the general magnetic index, $S_{ph}$, the red dot-dashed line corresponds to the mean magnetic index, $S_{ph,k}$.}
\label{Fig:sph1}
\end{center}
\end{figure}

As an example, Fig.~\ref{Fig:sph1} shows the light curve of the star KIC~6448798 from Q0 to Q14 that spans more than 1200 days (top left panel). The rotational period measured for this star is  $P_{\rm{rot}}=6.44 \pm 0.56$~days. The following panels show the evolution of the magnetic index  $S_{ph,k}$ calculated for different values of the factor $k$ (1, 3, 5, 10, and 30) as a function of time. 
The black dashed line represents the value of the standard deviation over the entire time series, $S_{ph}$, while the red dot-dashed line corresponds to the mean value of the standard deviations, $\langle S_{ph,k} \rangle$, calculated for each $k$.  The mean value instead of the median value is used because the median can underestimate the activity level if there is an on-going magnetic activity cycle. \citet{2014JSWSC...4A..15M} showed that the value of $\langle S_{ph,k} \rangle$ is slightly lower than $S_{ph}$, and the difference is lowest for a given value of $k$. They showed that a value of $5 \times P_{rot}$  describes the magnetic temporal evolution of stars reasonably well and also gives a correct value of general activity index. This was also clearly demonstrated in for the Sun with the photometric VIRGO/SPM observations.
A photometric magnetic activity level like this was recently used by \citet{2014A&A...563A..84G} to study KIC~8561221 and for a sample of  a few dozens of F \citep{2014A&A...562A.124M} and M~stars \citep{2014JSWSC...4A..15M}.
\begin{figure}[t]
\begin{center}
\includegraphics[width=0.51\textwidth]{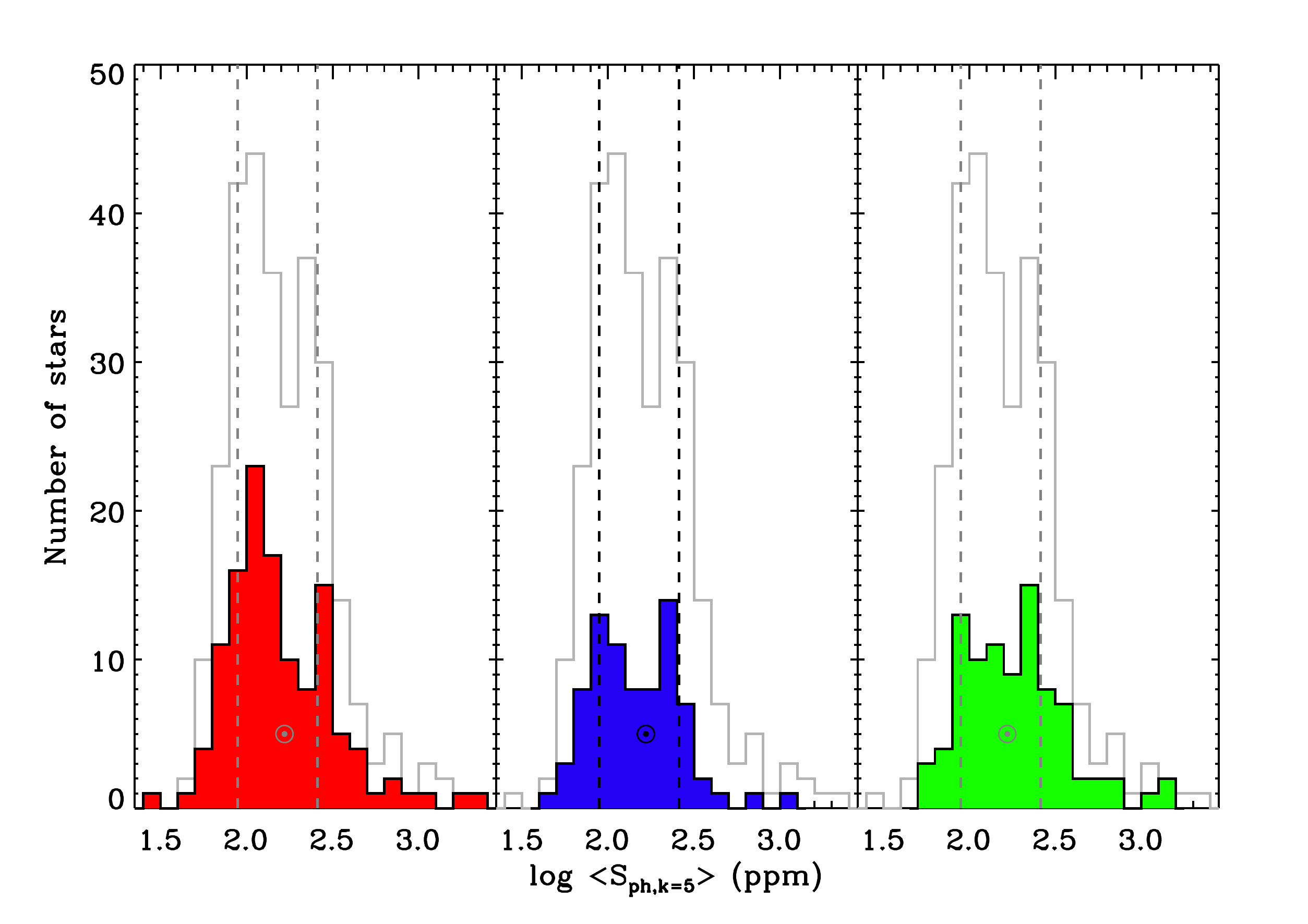}
\caption{Histograms of the extracted photometric magnetic activity level, $\log \langle S_{ph,k=5} \rangle$ for the full sample of stars for which we have retrieved a $P_{\rm{rot}}$  (grey), hot (red), dwarf (blue), and subgiant stars (green) as defined in Fig.~\ref{Fig:HR}. For reference, the activity of the Sun (166.1~$\pm$~2.6~ppm) is also shown. The vertical dashed lines enclose the solar magnetic-activity range (from 89 to 258.5 ppm). The solar values are colour-coded as in Fig.~\ref{Fig:histo2}.}
\label{Fig:histo_sph}
\end{center}
\end{figure}

Figure~\ref{Fig:histo_sph} shows the distribution of $\langle  S_{ph,k} \rangle$ for $5 \times P_{rot}$ for the 310 stars for which a rotation period $P_{\rm{rot}}$ was  measured (Sect.~\ref{Sec:rot}). The figure is colour-coded following the three categories defined in the previous section: hot stars, dwarfs, and subgiant stars. The distribution of the activity levels appear to be similar for the three categories of stars. Most of the stars in the sample have an  $\langle  S_{ph,k=5} \rangle$ located between the minimum and maximum solar magnetic activity of 89 $\pm$1.5 and 258.5 $\pm$ 3.5 ppm \citep{2014A&A...562A.124M}. In particular, of a total of 78 dwarf stars, 48 (61.5\%) lie between the limits of the solar activity, 19 (24.4\%) are more active, and 11 (14.1\%) are less active. The distribution of the hot
stars is skewed toward higher values than the distribution of
the dwarfs and subgiants. 

To place this study in a wider context, it is important to recall that the sample of stars used in this work were chosen because pulsations were measured. This means that because magnetic activity inhibits stellar pulsations \citep[e.g.][]{2010Sci...329.1032G,2011ApJ...732L...5C} and the amplitudes of the modes increase with age, our set of stars is biased towards aged stars (see Fig.\ref{Fig:HR}) with moderate magnetic activity.

\begin{figure}[!htb]
\begin{center}
\includegraphics[width=0.5\textwidth]{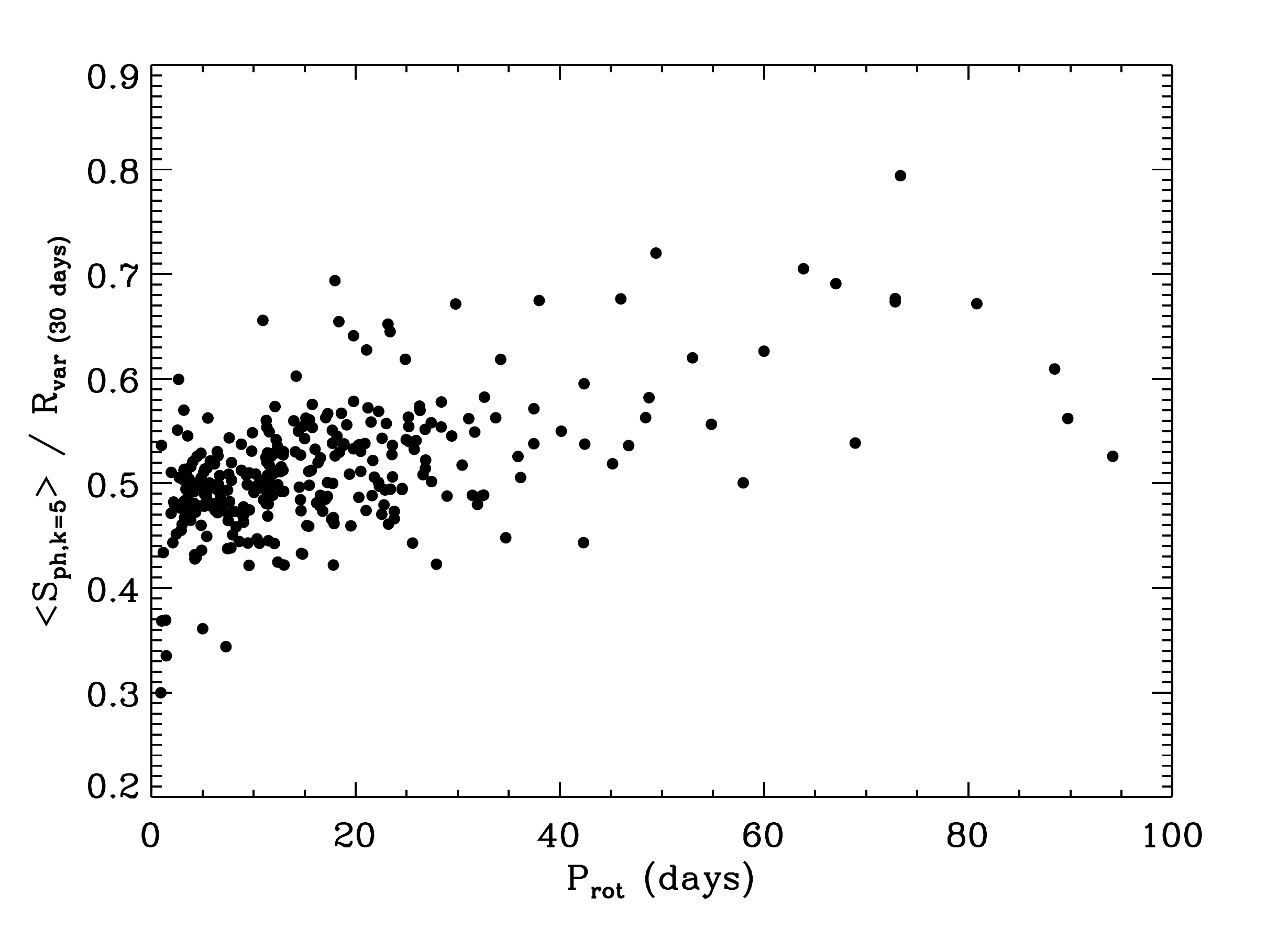}
\caption{Ratio between the measured photometric magnetic activity level, $\langle S_{ph,k=5} \rangle$, and the {\it range}, $R_{var}$(30 days) (Basri et al., 2013) as a function of the rotation period $P_{\rm{rot}}$  for the 310 solar-like pulsating stars observed by {\it Kepler}.}
\label{Fig:ratio_prot}
\end{center}
\end{figure}

The variability index proposed by \citet{2013ApJ...769...37B} was calculated on segments of 30 days of data, which naturally introduces a bias towards slow rotators. To estimate this bias, we calculated the {\it range}, $R_{var}$(30~days), as in \citet{2013ApJ...769...37B} for the 310 stars. Figure~\ref{Fig:ratio_prot} shows the ratio between the magnetic activity level, $\langle  S_{ph,k=5} \rangle$, and  $R_{var}$(30 days) as a function of the rotation period $P_{\rm{rot}}$.    A bias is clearly introduced for stars with a slower rotation period, $R_{var}$(30~days), underestimating their activity level by about 30\% compared with stars with a fast rotation period within the same sample.

\section{Discussion}
\label{Sec:disc}

We first compare our results with those obtained by \citet{2014arXiv1402.5694M}, who determined the $P_{\rm{rot}}$ of many main-sequence stars -- without binaries -- by studying the autocorrelation function of their PDC lightcurves. Their results were divided into two categories: periodic, when they had a reliable $P_{\rm{rot}}$, and non-periodic, when no $P_{\rm{rot}}$ could be obtained or when they found that the obtained $P_{\rm{rot}}$ was unreliable. Of our sample of reliable $P_{\rm{rot}}$, 114 stars are in common with their periodic sample. For these stars, both results agree
well within one sigma with the exception of  KIC~4931390, for which they derive a very short period of 0.57 days, while we find a much longer rotation period of 7.45 days. Moreover, we derived a reliable $P_{\rm{rot}}$ for 155
stars that they classified as non-periodic. For 129 of these 155 stars, their method still derives an unreliable rotation period. For these, the $P_{\rm{rot}}$ of 102 stars agree with our results within $2\sigma$, 17 stars agree with twice our $P_{\rm{rot}}$, and 1 star agrees with half our $P_{\rm{rot}}$. All of them are within $2\sigma$. This leaves only 9 stars for which the two analyses disagree. Because our methodology includes a visual inspection of every single target, the periods we derived for these stars are reliable. Moreover, when we did not derive a rotation period for a star, no rotation period was obtained by \citet{2014arXiv1402.5694M}.


\begin{figure*}[!ht]
\begin{center}
\includegraphics[width=0.86\textwidth]{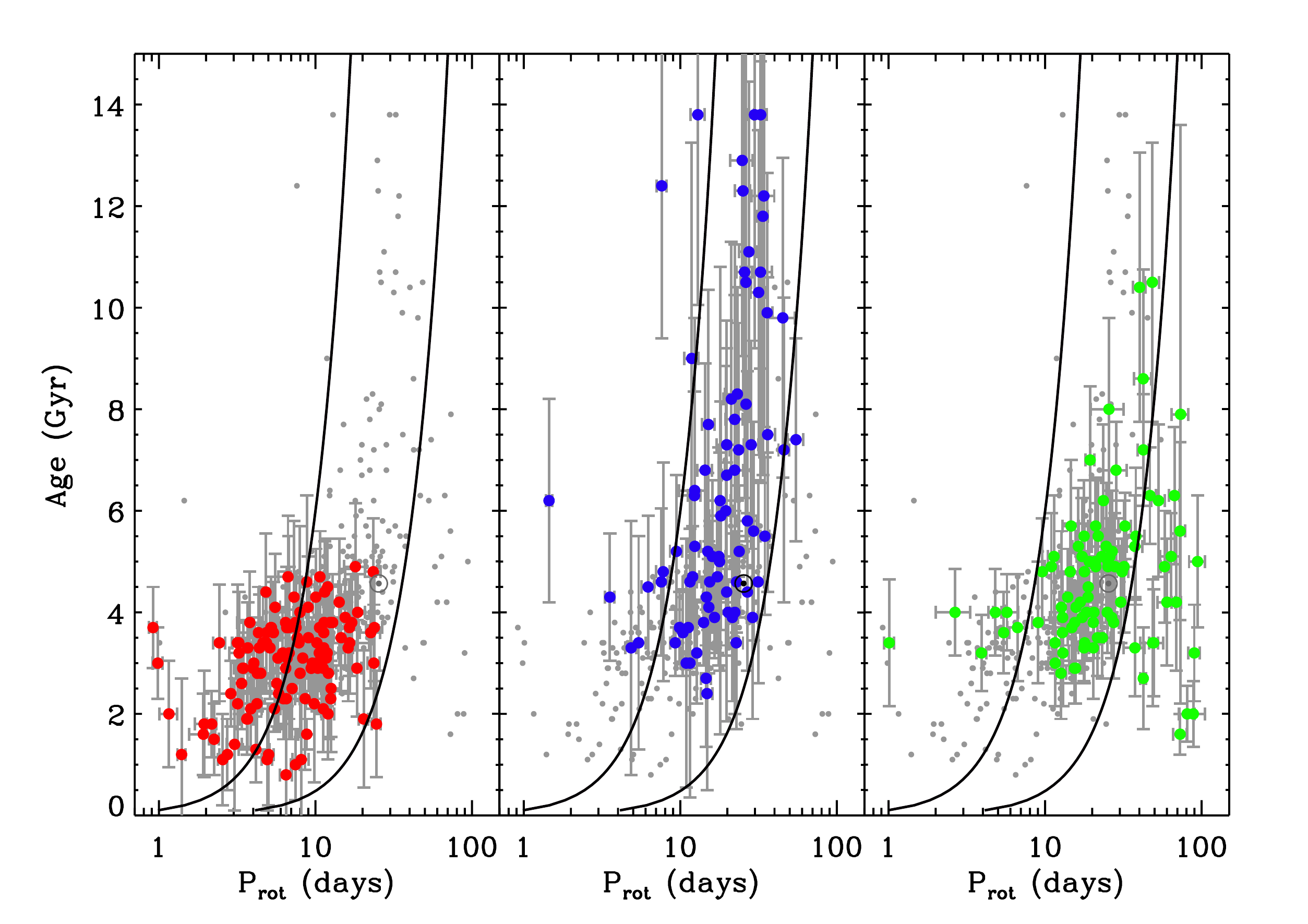}\\
\caption{Rotation periods measured in the this work as a function of {grid-modelling asteroseismic} ages taken from \citet{2014ApJS..210....1C}. Stars have been divided into hot (red), dwarfs (blue) and subgiants (green) as defined in Fig.~\ref{Fig:HR}. The solid black curves represent the period-age relationships from \citet{mamajek2008}, plotted for B-V =0.5  and B-V = 0.9, corresponding to late-F to early-K spectral types. The position of the Sun in the diagram is indicated by the $\odot$ symbol and colour-coded as in Fig.~\ref{Fig:histo2}. }
\label{Fig:gyro}
\end{center}
\end{figure*}

We then compared our results with the results obtained by \citet{2013MNRAS.433.3227K}, who determined the rotational periods, $P_{\rm{rot}}$, from the periodograms of the PDC \emph{Kepler} light curves, as well as the chromospheric activity $S$ indexes from spectroscopic observations for a small subset of the targets analysed in this paper.  In this sub-sample, \citet{2013MNRAS.433.3227K}  measured $P_{\rm{rot}}$ for ten stars. In eight stars both results agree within $1\sigma$. For KIC~8694723, we were unable to extract a reliable $P_{\rm{rot}}$ (see Table~4), while \citet{2013MNRAS.433.3227K} measured a rotation period of $7.5\pm0.2$~days. Finally, our results disagree for KIC~4914923. We found a rotation period of 20.49 $\pm$ 2.82 days, while they found 8.1 $\pm$ 0.4 days. However, \citet{2013MNRAS.433.3227K} limited their analysis of the periodograms to between 0 and 20 days. Thus, any $P_{\rm{rot}}$ longer than 20 days could not be detected. A close visual inspection of our GWPS and ACF analyses from the PDC-MAP and KADACS data sets shows a prominent and clear signature of the rotation around 20 days. However, a smaller peak around 8 days is also visible, but its signature is much fainter and might be caused by the third harmonic of $P_{\rm{rot}}$.


Among the stars with measured reliable periods we found 15 KOIs with three confirmed systems and one false positive. The extracted $P_{\rm{rot}}$ for two of them is a multiple of the transit period. All of them are single systems to the best of our knowledge. By studying  $P_{\rm{rot}}$  as a function of the orbital period of the innermost planet, we found that there are no close-in planets around fast rotators in this small sample of asteroseismic targets, as pointed out by \citet{2013ApJ...775L..11M}. The study of the KOI light curves requires a specific analysis to properly remove the transit signature without affecting the rotation modulation.

\subsection{Towards asterosismically calibrated age-rotation relations}
\label{sec:gyro}

The set of stars used in this analysis has the potential to provide calibration points for empirical gyrochronology relationships for field stars. We examined the relationship between surface rotation periods and ages by combining the periods provided in this work and the ages determined asteroseismically by  \citet{2012ApJ...749..152M}, \citet{2014ApJS..214...27M}, and \citet{2014ApJS..210....1C}. This is a first attempt to do so with a large asteroseismic sample. However, we recall that this set is biased by the necessity to detect acoustic modes in the stars and most of the stellar ages are determined from grid-modelling techniques using only  general seismic parameters  \citep[for further details see][]{2014ApJS..210....1C}.

First, it is important to consider our theoretical expectations regarding the relationship between rotation period and age. Cool main-sequence stars undergo magnetic braking and spin down as a function of time
\citep{1972ApJ...171..565S} and are therefore expected to display a correlation between period and age. Hot stars ($T_{\rm{eff}} >$ 6250 K), however, should not, since their thin convective envelopes and presumably weak dynamos result in very little magnetic braking. Subgiants, whose rotation rates are modified by the addition of substantial envelope expansion, are expected to display a relationship between period and age, but of a different form than that observed on the main sequence. 
We therefore divided our sample into three groups based on the stellar parameters in \citet{2014ApJS..210....1C}: cool main-sequence dwarfs (blue, $T_{\rm{eff}} <$ 6250 K, $\log g >$ 4.0), hot stars (red, $T_{\rm{eff}} >$ 6250 K), and subgiants (green, $T_{\rm{eff}} >$ 6250K, $\log g <$ 4.0).Ê

We show in Fig.~\ref{Fig:gyro} the grid-modelling asteroseismic ages from \citet{2014ApJS..210....1C} plotted against the $P_{\rm{rot}}$ from this work. These three subsets populate the period-age diagram in a manner consistent with the theoretical predictions of \citet{2013ApJ...776...67V} and demonstrate that care must be taken in interpreting the rotation periods of stars in such a sample.
 The curves represent the period-age relationships from \citet{mamajek2008} for B-V =0.5Ê and B-V = 0.9, corresponding to roughly 0.8 and 1.2 $M_{\odot}$. Stars within this mass range that obey the gyrochronology relationships should fall between these two curves, and the cool-star sample displays just this behavior (see middle panel in Fig.~\ref{Fig:gyro}). Hot stars, however, do not:Ê a many of the hot stars lie at systematically faster rotation periods and older ages than gyrochronology would have predicted. These stars do not spin down as a function of time because of weak magnetized stellar winds and therefore their period and age are
only weakly related (see left panel in Fig.~\ref{Fig:gyro}). Likewise, many of the subgiants fall outside of the gyrochronological period-age relationships, again because their rotational history is different from that of a cool main-sequence dwarf. They expand on the subgiant branch and continue to lose angular momentum through winds and as a result can populate the young, slowly rotating quadrant of this diagram (see right panel in Fig.~\ref{Fig:gyro}). Moreover, certain period regimes ($\sim$~20 days, for example) contain a mixture of dwarfs, hot stars, and subgiants, all of which have different relationships between period and age.Ê
  \begin{table}[!htb]
  \begin{center}
    \caption{\label{tbl_gyro} Linear fits $\log{P_{\rm{rot}}} = n \log(t) + c$, to the whole stellar sample as well as the three subsets. The total number of stars per category and those with $P_{\rm{rot}} > 5$ are also listed.}
    \begin{tabular}{lcccc}
      \hline \hline
      Category & $\#$  & $\#$  $P_{\rm{rot}} > 5$  & n & c \\
      \hline
      Whole sample & 289 & 243 & 3.87 $\pm$ 0.41 & -1.25 $\pm$ 0.27\\
      \hline
      Hot  & 122 & 83 & 5.80 $\pm$ 2.74 & -2.03 $\pm$ 2.10\\
      Cool MS dwarfs & 78 & 75 & 2.19 $\pm$ 0.38 & -0.42 $\pm$ 0.30\\
      Subgiants & 89 & 85 & 8.32 $\pm$ 2.89 & -4.23 $\pm$ 2.35\\
      \hline
    \end{tabular}
  \end{center}
\end{table}

 In empirical gyrochronology relations rotation is typically treated as a function of color (or mass) and age. As is evident in Fig.~\ref{Fig:gyro}, different populations in the HR diagram have very different relationships between rotation and age, as expected from stellar evolution theory. To illustrate this point, we performed a linear fit of the form $\log{P_{\rm{rot}}} = n \log(t) + c$,  where $P_{\rm{rot}}$ is given in days and $t$  in Gyr.
The results of the fit to the whole sample of stars and to each of the three subsets are summarised in Table~\ref{tbl_gyro}. We restricted the analysis to objects with periods of five days or longer to avoid any potential contamination from synchronized binaries, or very young objects for which the period-age relationships are less reliable \citep[see][]{2014ApJ...780..159E}.

As expected, the results confirm our theoretical expectations. When the whole sample is fitted, including all evolved, dwarf, and hot stars, the power-law slope of the relation is $\sim 3.9$: an unreasonable value, derived from an unreasonably mixed population of stars. Similarly unreasonable results are obtained for the hot star and subgiant samples because the gyrochronology relations are not applicable for these sets of stars. Finally, the set of cool main-sequence dwarfs has the closest slope $n$ to other values found in the literature, although it is still too high: the original \citet{1972ApJ...171..565S} relation  with $n = 0.51$, \citet{2007ApJ...669.1167B}  with $n = 0.5189 \pm 0.0070$, and \citet{mamajek2008} with $n = 0.566 \pm 0.008$.  
This slope difference can be caused by many factors.  It might be a reflection of the mass dependence of rotation, the admixture of main-sequence and subgiant stars even in our sample, metallicity effects, or the relatively large age errors associated with photometric metallicities.
 
 \begin{figure}[!htbp]
\begin{center}
\includegraphics[width=0.48\textwidth]{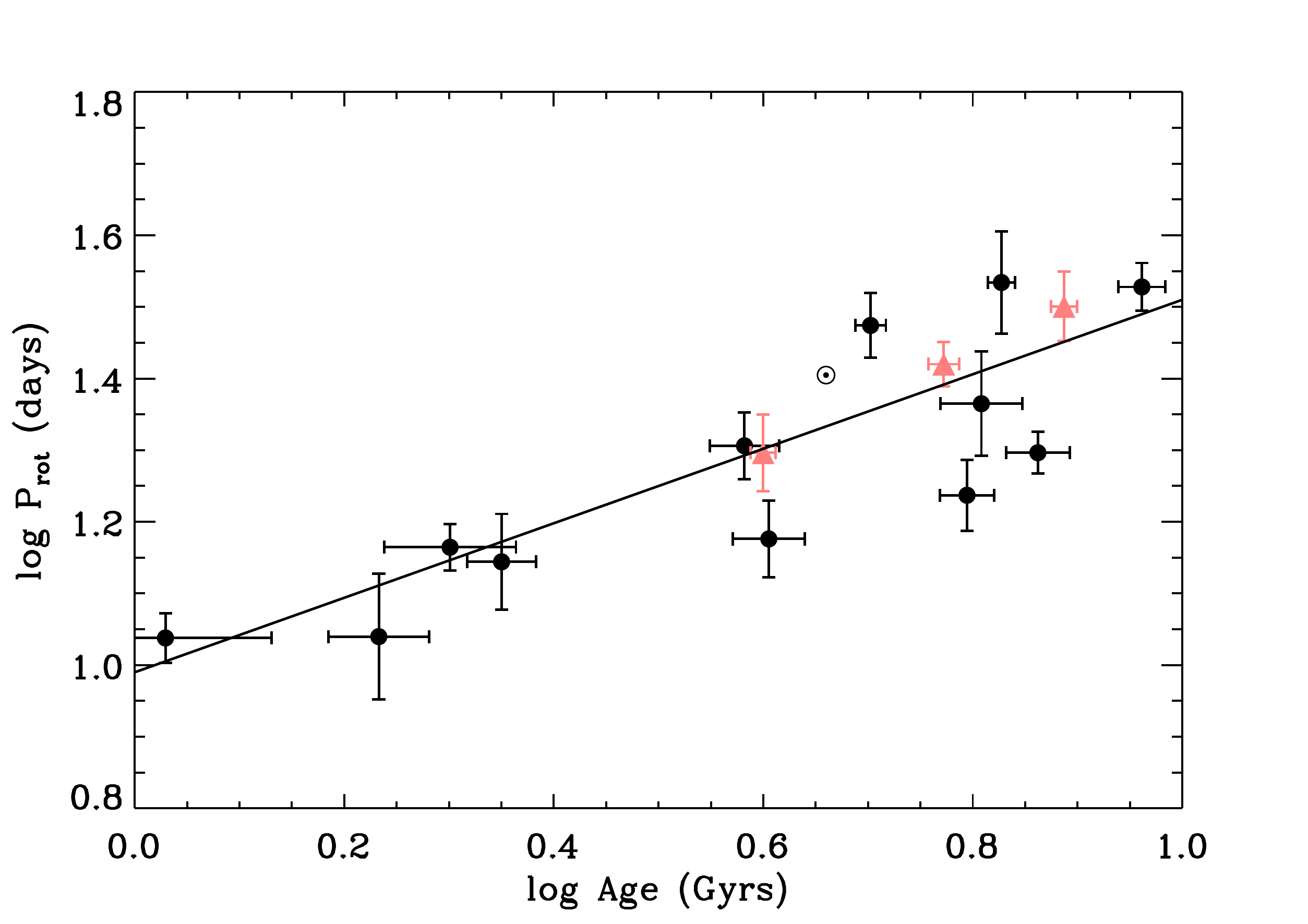}
\caption{Rotation periods, $P_{\rm{rot}}$, as a function of asteroseismic ages computed by individual modelling of stars in a log-log space. These ages are given by \citet{2014ApJS..214...27M} for 12 stars (black circles) and \citet{2012ApJ...749..152M} for 3 other stars (pink triangles). The solid line corresponds to a linear fit including the Sun. 
}
\label{Fig:rot_Met}
\end{center}
\end{figure}

To proceed, we selected a smaller sample of cool dwarfs for which we have precise ages derived from the modelling of stars where individual p-mode frequencies and spectroscopic effective temperatures and metalicities were used to find the best-fit models.
We thus took 11 stars analysed by \citet{2014ApJS..214...27M}, KIC~6116048, KIC~7871531, KIC~8006161, KIC~8228742, KIC~9098294, KIC~9139151, KIC~9955598, KIC~10454113, KIC~10644253, KIC~11244118, and KIC~12258514,  as well as 3 more stars from \citet{2012ApJ...749..152M}, KIC~3656476, KIC~5184732, and KIC~7680114. We added to this sample KIC~3427720, which  fits the same criteria and is registered as a star in a mutliple system by SIMBAD only because it is one of two components of a widely separated visual binary. Its light
curve is therefore not polluted by its companion. This subset represents the best-characterised stars so far to test gyrochronology. We also added the Sun to this  sample and considered a $P_{\rm{rot,\odot}}$ of 25.4 days with an adopted error of 10\% (to take into account differential rotation) and an age of 4.570 $\pm$ 0.007 years \citep{1995RvMP...67..781B}. We repeated the linear fit in the $\log(t) - \log(P_{\rm{rot}})$ space (see Fig.~\ref{Fig:rot_Met}):
\begin{equation}
\log{P_{\rm{rot}}} = (0.52 \pm 0.06) \log(t) + (0.99 \pm 0.04).
\end{equation}

\begin{figure*}[!htbp]
\begin{center}
\includegraphics[width=0.86\textwidth]{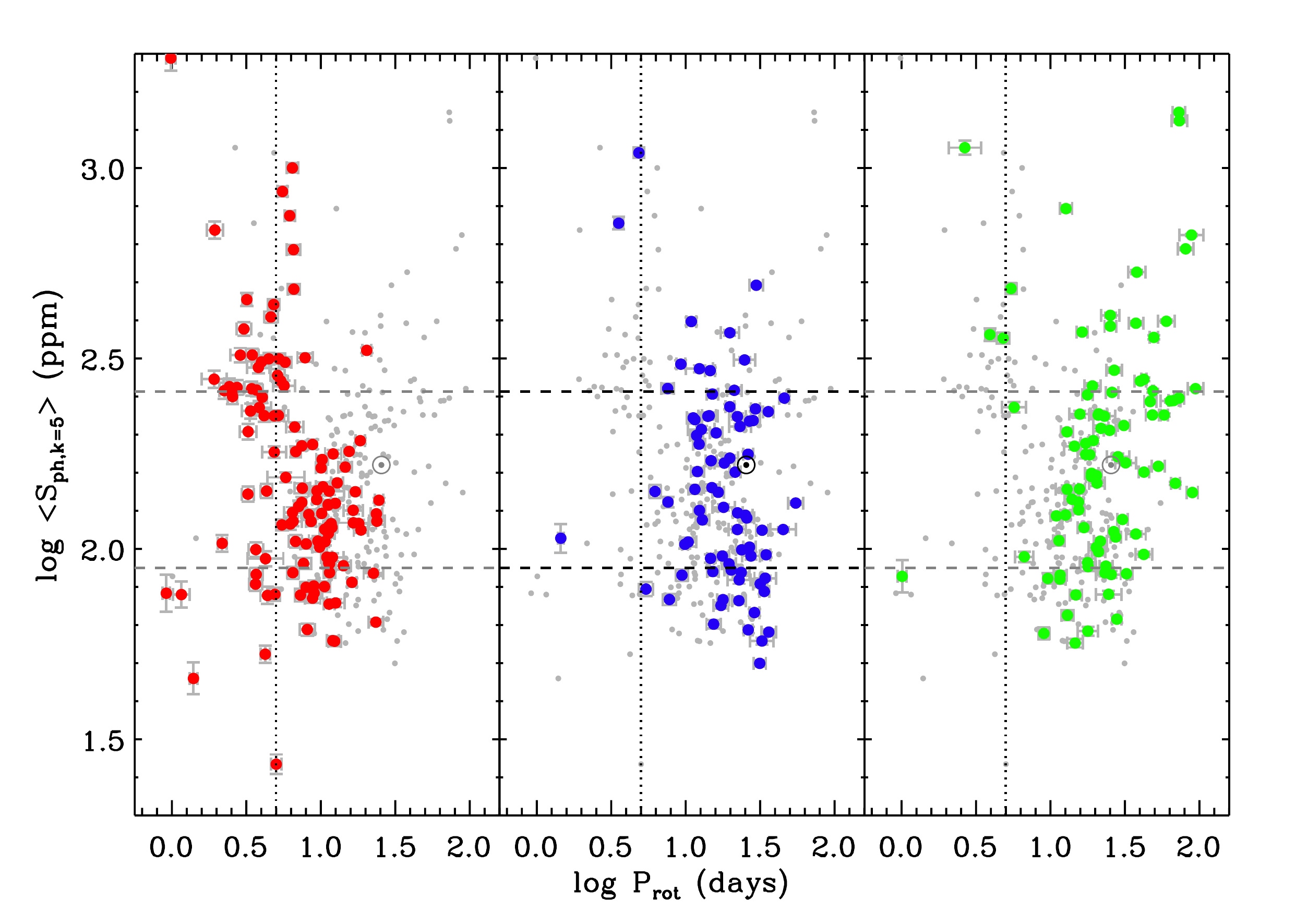}
\caption{Photometric magnetic activity level, $\langle S_{ph,k=5} \rangle$, as a function of the rotation period, $P_{\rm{rot}}$, for the 310 solar-like pulsating stars observed by {\it Kepler} for which the rotation period was successfully measured (grey). Stars have been divided into hot (red), dwarf (blue), and subgiants (green) as defined in Fig.~\ref{Fig:HR}. The vertical black dotted lines mark the limit of $P_{\rm{rot}}$ = 5 days. The position of the Sun, $\langle S_{ph,k=5} \rangle$ =166.1 $\pm$ 2.6 ppm and $P_{\rm{rot}}$=25.4~days  is indicated. The top and bottom dashed horizontal lines represent the corresponding solar magnetic activity level at maximum and minimum of the 11-year solar cycle. The solar values are colour-coded as in Fig.~\ref{Fig:histo2}.}
\label{Fig:sph_vs_prot}
\end{center}
\end{figure*}

In this case, the slope agrees very well with previously published slopes including the value of $n = 0.41 \pm 0.03$  derived in \citet{2014ApJS..214...27M} using a similar data set. From the \citet{mamajek2008} calibration, one expects the intercept of the period-age relation for the typical star in our sample with B-V  = +0.58 ($\sim$ 1.05 $M_{\odot}$) to be $0.96 \pm0.05$, or $1.00 \pm 0.04$ from the Barnes calibration, both of which are consistent with our result. 
A more in-depth analysis of the gyrochronological consistency and value of this asteroseismic data set is reserved for later papers; for now, it is sufficient to emphasize the fact that stellar populations, adequate sample selection, and precise asteroseismic ages properly constrained from spectroscopy are important when deriving period-age relationships. Careful angular momentum evolution modelling, including the effects of structural change and metallicity, will also be important for interpreting the observations.

Future work will investigate the behavior of the gyrochronology relations at late times in detail. Our work here highlights the importance of stellar populations in interpretating the rotation periods.


\subsection{Activity-rotation and age-activity relations}

Figure~\ref{Fig:sph_vs_prot} shows on a log-log scale the measured photometric magnetic activity proxy.  The three groups of stars defined in Sect.~\ref{sec:gyro} are represented in separate panels with the same colour-coding as used in Fig.~\ref{Fig:gyro}. The solar value is also represented, while the horizontal dashed lines represent the corresponding solar magnetic activity level at minimum and maximum of the 11-year solar cycle \citep{2014A&A...562A.124M}. Some of the stars with a high $\langle  S_{ph,k=5} \rangle$ and a low $P_{\rm{rot}}$ might be possible multiple systems because they have very stable ACFs.

\begin{figure*}[!htbt]
\begin{center}
\includegraphics[width=0.86\textwidth]{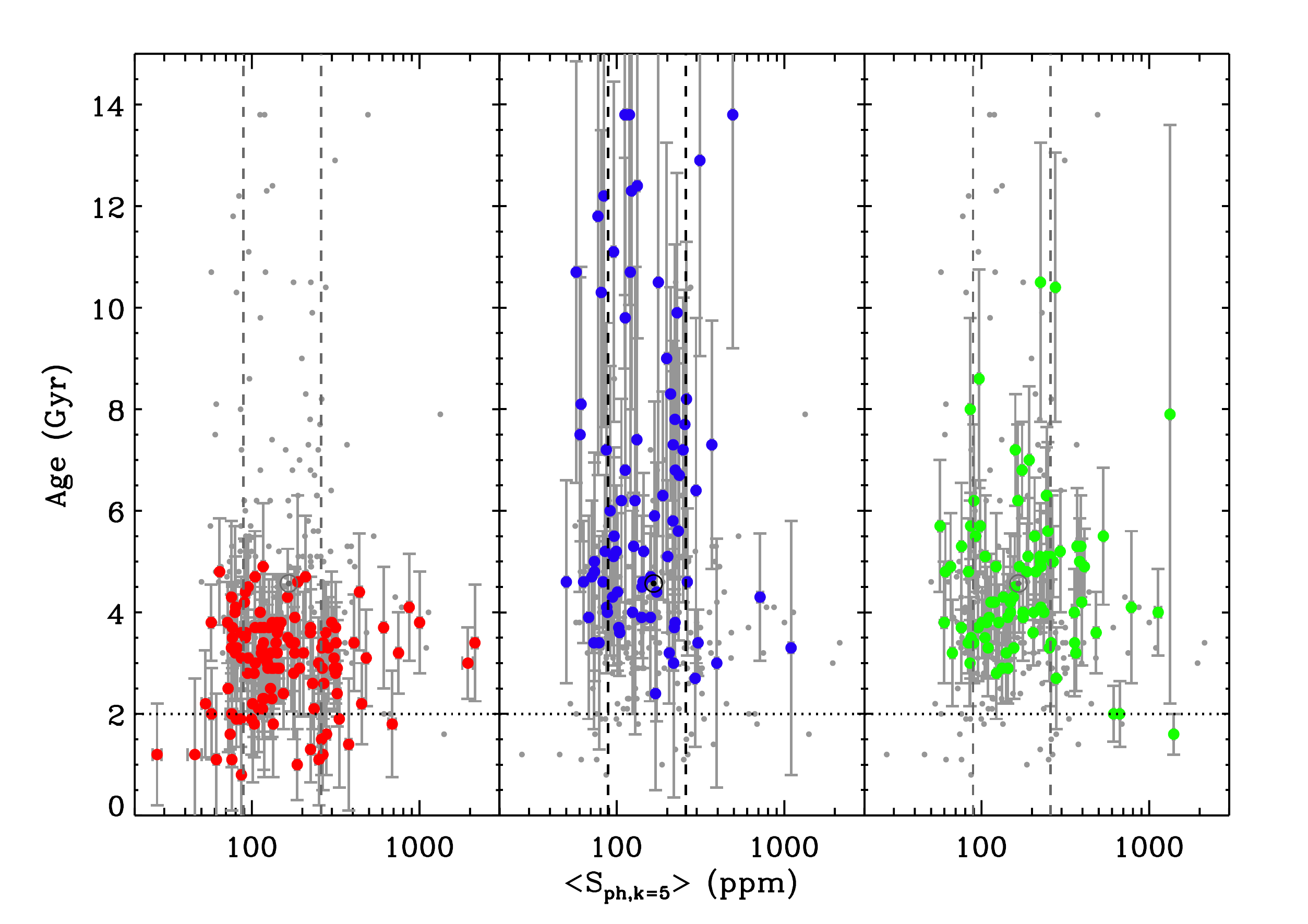}
\caption{Photometric magnetic activity level, $\langle S_{ph,k=5} \rangle$,  as a function of  grid-modelling asteroseismic ages (in grey) taken from \citet{2014ApJS..210....1C}. Stars have been divided into hot (red), dwarf (blue), and subgiants (green) as defined in Fig.~\ref{Fig:HR}. Only stars with $P_{\rm{rot}} \geq 5$ days are shown. Because of the size of the error bars, they are only plotted for each category of stars. The solar value of $\langle S_{ph,k=5} \rangle=$  166.1~$\pm$~2.6~ppm for an age of 4.57~Gyr  is represented with the solar symbol and the same colour code as in Fig.~\ref{Fig:histo2}. The left and right dashed vertical lines correspond to the solar magnetic activity level at minimum and maximum of the 11-year solar cycle. The horizontal dotted lines mark the 2~Gyr limit.}
\label{Fig:age_vs_sph}
\end{center}
\end{figure*}

For each subsample, we determined the correlations between $\log P_{\rm{rot}}$ and $\log S_{ph,k=5}$. We first considered stars with $P_{\rm{rot}}>5$~d for the same reasons as explained in Sect.~\ref{sec:gyro}. Although it has been demonstrated that evolved binaries with long rotational periods can also show high magnetic activity levels \citep[e.g.][]{1993A&AS..100..173S,1995PASP..107..503M,2000A&AS..146..103M}, it is beyond the scope of this paper to analyse them in detail. In a future work we will study the rotation and magnetism of the \emph{Kepler} binaries.

 For hot stars, cool dwarf stars, and subgiants, the correlations are -31\%, -15\%, and 43\%, respectively. By restricting the sample to $P_{\rm{rot}}>10$~d, the correlations become 12\%, -27\%, and 49\%. The correlations are weak or very weak. For hot stars, the slight anti-correlation in the first sample was only due to a few stars with high $S_{ph,k=5}$ and $P_{\rm{rot}}$ between five and ten days, clearly visible in Fig.~\ref{Fig:age_vs_sph}. The highest correlation is found for the subgiants that could follow an activity-rotation relation for older stars, as pointed out for example by \citet{2008ssma.book.....S}.


To understand this loose correlation, we need to note that the dispersion of $S_{ph,k=5}$ in the sample is similar to the range spanned by the Sun during the magnetic activity cycle. As a consequence, $S_{ph,k=5}$  reflects the average photospheric magnetic activity during any on-going activity cycle longer than the current length of the {\it Kepler} observations. Therefore, this might help to explain the high spread in the results \citep[e.g.][]{2010A&A...514A..97L}. A ground-based continuous monitoring of these stars for longer times is necessary to properly analyse any  correlation between rotation periods and magnetic activity in this sample of stars. Another source of dispersion might be the wide range of masses and metallicities of the stars in our sample. Finally, the stellar inclination angle has a direct impact on the determination of  $S_{ph,k=5}$  , which might likewise contribute to the spread in the results. For all these reasons, we are unable to obtain a reliable relation between the surface rotation rate and the photospheric magnetic activity.

Figure~\ref{Fig:age_vs_sph} shows the age-activity relation for the set of stars with measured $P_{\rm{rot}}$, taking the ages from \citet{2014ApJS..210....1C}. Although the dispersion is high here as well for the same reasons as before, we can investigate the behavior of young stars below 2 Gyr. In previous works, the age-activity relation derived from the chromospheric S-index showed a rapid decay of chromospheric activity up to 2~Gyr, and no decay above that \citep{2013A&A...551L...8P}. Unfortunately, most of our stars are concentrated in a bulk between 3 and 5~Gyr, and only $\sim$4\% are younger than 2~Gyr, which makes it a comparison with the observations of \citet{2013A&A...551L...8P} difficult. Nevertheless, it is worth noting that while the photometric magnetic activity levels $\langle  S_{ph,k=5} \rangle$ are mostly in the range between 60 and 300~ppm, some of the hot stars and subgiants  show high activity levels around 3~Gyr and younger. In contrast, these high activity levels are not observed in dwarf stars. This might be a selection bias: our stars were selected because they had acoustic modes and we know that activity reduces their amplitudes  \citep{2010Sci...329.1032G,2011ApJ...732L...5C,2014ApJ...783..123C}. Nevertheless, hot stars and subgiants with high activity levels younger then 3~Gyr might belong to the stars located at the end of the activity decay observed by \citet{2013A&A...551L...8P}.

\section{Conclusions}
\label{Conclu}
We have analysed a homogeneous set of 540 main-sequence and subgiant pulsating stars presented in \citet{2014ApJS..210....1C} and extracted reliable surface rotation periods and photometric activity indexes for 310 stars. To do so, we combined two detection methods (GWPS and ACF) with two ways of preparing the light curves (PDC-MAP and KADACS). Special care was taken to properly identify all the binaries in the sample in the bibliography.

These stars were divided into three different categories, hot stars ($T_{\rm{eff}} >$ 6250 K), cool main-sequence dwarfs ($T_{\rm{eff}} \leq$ 6250 K, $\log g >$ 4.0), and subgiants (green, $T_{\rm{eff}} \leq$ 6250K, $\log g \leq$ 4.0). As expected, the hotter stars spin faster than the cool main-sequence dwarfs because their thin convective envelopes and presumably weak dynamos result in very weak magnetic braking. Subgiants can have periods of $\sim$10-100 days, depending on the main-sequence temperature and degree of expansion that the star has undergone on the subgiant branch. 

A total of 15 KOIs were in our sample. We failed to find any close-in planet around fast-rotating stars, confirming the results obtained by  \citet{2013ApJ...775L..11M}.

We found important differences in the rotation-age relationship between hot dwarfs, cool dwarfs, and subgiants. These differences highlight the importance of population effects for interpreting gyrochronology relationships. A subset of the data in which we have very precise age estimates from the detailed analysis of individual frequencies and spectroscopic constraints has a slope different from that of the entire sample and consistent with expectations from the literature. As soon as more precise asteroseismic ages (based on individual oscillation frequencies and spectroscopic observations) are determined for other stars in this sample, it will be possible to expand the gyrochonology seismic analysis to more  \emph{Kepler} field stars.

 We found that the photometric magnetic activity $\langle S_{ph,k=5} \rangle$ for most of the solar-like pulsating stars in our sample is similar to that of the Sun during its magnetic activity cycle. Indeed, 61.5\% of the dwarfs have values similar to those of
the Sun. However, the high dispersion found in our results might reflect that we did not cover the full magnetic activity cycle of many of the stars in our sample. Other factors, such as the unknown stellar inclination axis, will also contribute to this dispersion. Therefore, we for stars similar to the Sun, $\langle S_{ph,k=5} \rangle$ is probably a good indicator of the magnetic activity of the star during the observed time. However, because \emph{Kepler} has ``only'' observed during four years so far, the variability of the star might not be representative of the average stellar magnetic activity during its full stellar magnetic cycle, and hence, we were unable to extract any reliable activity-rotation or age-activity relations. Further studies will be necessary to extract a subset of stars for which at least a full magnetic cycle has been observed and then be able to properly establish these relations.




\begin{acknowledgements} 
The authors wish to thank the entire \emph{Kepler} team, without whom these results would not be possible. Funding for this Discovery mission is provided by NASA's Science Mission Directorate. 
We also thank all funding councils and agencies that have supported the activities of KASC Working Group 1, as well as the International Space Science Institute (ISSI).  This research was supported in part by the National Science Foundation under Grant No. NSF PHY05-51164. The research leading to these results has received funding from the European CommunityÕs Seventh Framework Programme ([FP7/2007-2013]) under grant agreement no. 312844 (SPACEINN), under grant agreement no. 269194 (IRSES/ASK), under the ERC grant agreement no. 227224 (PROSPERITY), and from the Research Council of the KU Leuven under grant agreement GOA/2013/012. SB is supported by the Foundation for Fundamental Research on Matter (FOM), which is part of the Netherlands Organisation for Scientific Research (NWO). DS, RAG, SM, and TC received funding from the CNES GOLF and CoRoT grants at CEA. RAG also acknowledges the ANR (Agence Nationale de la Recherche, France) program IDEE (n¡ ANR-12-BS05-0008) ``Interaction Des \'Etoiles et des Exoplan\`etes''. The Danish National Research Foundation (Grant agreement no.: DNRF106). The research is supported by the ASTERISK project (ASTERoseismic Investigations with SONG and \emph{Kepler}) funded by the European Research Council (Grant agreement no.: 267864). MBN acknowledges research funding by Deutsche Forschungsgemeinschaft (DFG) under grant SFB 963/1 ``Astrophysical flow instabilities and turbulence'' (Project A18). This research has made use of the SIMBAD database, operated at CDS, Strasbourg, France.
\end{acknowledgements} 

\bibliographystyle{aa}

\begin{table*}
  \caption{\label{tbl_rot} Stars for which we derived a reliable rotation period $P_{\rm rot}$. KOIs are identified by a $^\dagger$ in the KID (\emph{Kepler} IDentification number. The column ``Type of detection'' explains whether $P_{\rm rot}$ has been determined automatically (A) or after a visual check (V). The column ``Source of $P_{\rm rot}$'' gives the set of data -- KADACS or PDC --  from which the final $P_{\rm rot}$ has been obtained. The $^\star$ indicates when the KADACS datasets were filtered at 100 days instead of the standard 30 days. This table is available in its entirety in a machine-readable form in the online journal. A portion is shown here for guidance
regarding its form and content.}
  \begin{tabular}{ccccccc}
\hline \hline
KID &  $P_{\rm rot}$ [days] & error $P_{\rm rot}$ [days] & $\langle S_{ph,k=5} \rangle$ & error $\langle S_{ph,k=5} \rangle$ & Type of detection & source of $P_{\rm rot}$ \\
\hline
1430163 & 4.16 & 0.92 & 223.79 & 8.18 & A & KADACS \\
1435467 & 6.68 & 0.89 & 208.99 & 5.81 & A & KADACS \\
2450729 & 53.00 & 5.32 & 164.74 & 1.78 & A & KADACS$^\star$ \\
2837475 & 3.68 & 0.36 & 85.68 & 3.46 & A & KADACS \\
2852862 & 10.13 & 0.63 & 123.90 & 3.33 & A & KADACS \\
2865774 & 6.22 & 0.56 & 141.49 & 5.49 & V & PDC \\
2998253 & 6.79 & 0.53 & 180.04 & 5.90 & A & KADACS \\
3112152 & 15.48 & 1.19 & 180.25 & 4.11 & V & KADACS \\
3123191 & 20.33 & 1.38 & 332.06 & 5.29 & A & KADACS \\
3223000 & 4.85 & 0.39 & 1096.52 & 34.88 & A & KADACS \\
3236382 & 3.66 & 0.42 & 99.56 & 4.40 & V & PDC \\
3241581 & 26.26 & 2.01 & 177.18 & 2.65 & V & KADACS \\
3344897 & 5.66 & 0.94 & 268.28 & 8.52 & A & KADACS \\
3424541 & 3.46 & 0.33 & 322.94 & 12.57 & A & KADACS \\
3430893 & 8.29 & 0.69 & 123.11 & 3.65 & A & KADACS \\
3632418$^\dagger$ & 12.89 & 1.38 & 133.50 & 2.71 & A & KADACS \\
3633847 & 10.96 & 1.03 & 122.10 & 3.01 & A & KADACS \\
3633889 & 4.23 & 0.30 & 52.90 & 2.71 & A & KADACS \\
3642422 & 48.73 & -8.19 & 259.91 & 2.92 & V & PDC \\
3643774 & 14.40 & 1.56 & 223.65 & 4.36 & A & KADACS \\
3656476 & 31.67 & 3.53 & 80.90 & 1.48 & A & KADACS \\
3657002 & 16.02 & 1.54 & 201.59 & 4.00 & A & KADACS \\
3661135 & 28.39 & 4.83 & 174.53 & 3.10 & A & KADACS \\
3733735 & 2.56 & 0.19 & 251.21 & 11.65 & A & KADACS \\
...\\
\hline

  \end{tabular}
\end{table*}

\begin{table*}
  \caption{\label{tbl_mult}  Table of the multiple stars in our sample. Columns are the same as in Table~\ref{tbl_rot}, except for the column ``Type'' in which the type of multiple star is given (EB: eclipsing binary, SB2: double-lined spectroscopic binary, Mult: multiple in SIMBAD database, Seis/Pol: seismic binary or pollution by a nearby red giant) and the column ``Reference',' which explains from which paper or website this classification comes from. Stars for which no rotation period was derived have $P_{\rm rot}=\langle S_{ph,k=5} \rangle=-1$. The star for which KID shows a $^\dagger$ is a KOI.}
  \begin{tabular}{ccccccc}
\hline \hline
KID &  $P_{\rm rot}$ [days] & error $P_{\rm rot}$ [days] & $\langle S_{ph,k=5} \rangle$ & error $\langle S_{ph,k=5} \rangle$ & Type & Reference \\
\hline
2010607 & 17.60 & 3.89 & 264.73 & 5.06 & EB & Kirk et al. 2013, in prep. \\
3427720 & 13.94 & 2.15 & 106.72 & 2.27 & Mult. & SIMBAD \\
4586099 & 10.76 & 1.94 & 137.62 & 3.25 & Mult. & SIMBAD \\
5021689 & -1.00 & 0.00 & -1.00 & 0.00 & Mult. & SIMBAD \\
7510397 & -1.00 & 0.00 & -1.00 & 0.00 & Mult. & SIMBAD \\
7938112 & 0.84 & 0.07 & 214.08 & 19.03 & Seis/Pol & This study \\
8360349 & -1.00 & 0.00 & -1.00 & 0.00 & Mult. & SIMBAD \\
8379927 & 16.99 & 1.35 & 1340.50 & 23.44 & SB2 & Griffin 2007 \\
9025370 & 13.31 & 1.30 & 171.69 & 3.63 & SB2 & Thygesen et al. 2014 \\
9139151 & 10.96 & 2.22 & 187.29 & 4.22 & Mult. & SIMBAD \\
9139163 & 6.10 & 0.47 & 71.23 & 2.32 & Mult. & SIMBAD \\
9693187 & -1.00 & 0.00 & -1.00 & 0.00 & SB2 & Thygesen et al. 2014 \\
9908400 & 17.12 & 1.26 & 530.12 & 10.29 & Mult. & SIMBAD \\
10124866 & 17.59 & 2.27 & 248.23 & 4.38 & Mult. & SIMBAD \\
11453915$^\dagger$ & 23.85 & 1.55 & 75.45 & 1.38 & EB & Kirk et al. 2013, in prep. \\
\hline
\hline
  \end{tabular}
\end{table*}

\begin{table*}
  \caption{\label{tbl_no_rot} Stars without a  sign of rotation in the light curve. The star for which KID shows a  $^\dagger$ is a KOI.}
  \begin{tabular}{c}
\hline \hline
KID \\
\hline
4915148 \\
4918355 \\
5603483 \\
5629080 \\
8179973$^\dagger$ \\
8547279 \\
\hline
  \end{tabular}
\end{table*}

\end{document}